\pdfoutput=1
\documentclass[a4paper,10pt,twocolumn]{article}

\usepackage[utf8]{inputenc}
\usepackage[T1]{fontenc}
\usepackage[english]{babel}
\usepackage{amsmath}
\usepackage{mathtools}
\usepackage{comment}
\usepackage{textcomp}
\usepackage[style=base]{caption}
\usepackage{bm}
\usepackage{url}
\usepackage{pdflscape}
\usepackage[margin=15mm,tmargin=2cm,bmargin=2.5cm]{geometry}

\usepackage{setspace}
\usepackage[bitstream-charter]{mathdesign}
\usepackage{sectsty}
\sectionfont{\fontsize{12}{15}\selectfont}
\subsectionfont{\fontsize{10}{13}\selectfont}

\usepackage{csquotes}
\usepackage[
    backend=biber,
    citestyle=numeric,
    bibstyle=phys,
    articletitle=false,
    biblabel=brackets,
    chaptertitle=false,
    pageranges=false,
    isbn=false,
    url=false,
    sorting=none
]{biblatex}
\usepackage[symbolpackage=l3draw, symbol=plos]{biblatex-ext-oa}
\DeclareFieldFormat{titlecase}{#1}
\bibliography{rsc}

\usepackage{overpic}
\usepackage[normalem]{ulem}

\newcommand{\mytitle}{Steering droplets on substrates using moving steps in wettability}
\newcommand{\authors}{Josua Grawitter and Holger Stark}

\renewcommand{\vec}[1]{\bm{#1}}
\newcommand{\tens}[1]{\mathsf{#1}}
\newcommand{\dif}{\,\mathrm{d}}

\usepackage[
    colorlinks=false,
    pdfborder={0 0 0},
    linktoc=all,
    pdfpagelabels,
    plainpages=false,
    bookmarks=true,
    hypertexnames=false, 
    pdftex,
    pdfauthor={\authors},
    pdftitle={\mytitle}
]{hyperref}
\usepackage{cleveref}

\newcommand*{\halurl}[1]{http://hal.archives-ouvertes.fr/#1}
\DeclareFieldFormat{eprint:hal}{%
  \ifhyperref
    {\href{\halurl{#1}}{hal:~\nolinkurl{#1}}}
    {hal:~\nolinkurl{#1}}}
\DeclareFieldAlias{eprint:HAL}{eprint:hal}

\DeclareOpenAccessEprintUrl{hal}{%
  http://hal.archives-ouvertes.fr/\thefield{eprint}}
\DeclareOpenAccessEprintAlias{HAL}{hal}

\newcommand*{\semschurl}[1]{https://api.semanticscholar.org/CorpusID:#1}
\DeclareFieldFormat{eprint:semanticscholar}{%
  \ifhyperref
    {\href{\semschurl{#1}}{S2CID:~\nolinkurl{#1}}}
    {S2CID:~\nolinkurl{#1}}}
\DeclareFieldAlias{eprint:SEMANTICSCHOLAR}{eprint:semanticscholar}

\DeclareOpenAccessEprintUrl{semanticscholar}{%
  https://api.semanticscholar.org/CorpusID:\thefield{eprint}}
\DeclareOpenAccessEprintAlias{SEMANTICSCHOLAR}{semanticscholar}

\newcommand*{\caltechurl}[1]{https://resolver.caltech.edu/CaltechAUTHORS:#1}
\DeclareFieldFormat{eprint:caltech}{%
  \ifhyperref
    {\href{\caltechurl{#1}}{CaltechAUTHORS:~\nolinkurl{#1}}}
    {CaltechAUTHORS:~\nolinkurl{#1}}}
\DeclareFieldAlias{eprint:CALTECH}{eprint:caltech}

\DeclareOpenAccessEprintUrl{caltech}{%
  https://resolver.caltech.edu/CaltechAUTHORS:\thefield{eprint}}
\DeclareOpenAccessEprintAlias{CALTECH}{caltech}

\usepackage{microtype}

\title{\mytitle}
\author{\authors}

\begin{document}
\renewcommand{\thefootnote}{\alph{footnote}}
\newgeometry{left=15mm,right=45mm}
\onecolumn
\noindent
{\huge \textbf{Steering droplets on substrates using\newline moving steps in wettability}$^\dag$}

{\large\onehalfspacing

\vspace{0.5cm}
\noindent \textit{
Josua Grawitter\footnotemark[1]\textsuperscript{,}\footnotemark[2] and Holger Stark\footnotemark[1]\textsuperscript{,}\footnotemark[3]
}
\hfill November~21, 2020

\footnotetext[1]{Institut für Theoretische Physik, Technische Universität Berlin, Hardenbergstr.~36, 10623 Berlin, Germany.}
\footnotetext[2]{E-mail:~\href{mailto:josua.grawitter@physik.tu-berlin.de}{\nolinkurl{josua.grawitter@physik.tu-berlin.de}}}
\footnotetext[3]{E-mail:~\href{mailto:holger.stark@tu-berlin.de}{\nolinkurl{holger.stark@tu-berlin.de}}}
\footnotetext{\hspace{-4pt}$^\dag$Electronic Supplementary Information (ESI) available: Their descriptions are collected in Appendix~\ref{appendix_esi}.}

\renewcommand{\thefootnote}{\arabic{footnote}}

\vspace{1cm}

\noindent\textbf{Abstract}\newline
Droplets move on substrates with a spatio-temporal wettability pattern as generated, for example, on light-switchable surfaces.
To study such cases,  we implement the boundary-element method to solve the governing Stokes equations for the fluid flow field inside and on the surface of a droplet and supplement it by the Cox-Voinov law for the dynamics of the contact line.
Our approach reproduces the relaxation of an axisymmetric droplet in experiments, which we initiate by instantaneously switching the uniform wettability of a substrate quantified by the equilibrium contact angle.
In a step profile of wettability the droplet moves towards higher wettability.
Using a feedback loop to keep the distance or offset between step and droplet center constant, induces a constant velocity with which the droplet surfes on the wettability step.
We analyze the velocity in terms of droplet offset and step width for typical wetting parameters.
Moving instead the wettability step with constant speed, we determine the maximally possible droplet velocities under various conditions.
The observed droplet speeds agree with the values from the feedback study for the same positive droplet offset.

} 


\restoregeometry
\twocolumn
\section{Introduction}
If a droplet can go uphill and move in response to light, where are the limits of its motility?
In their seminal experiment, Chaudhury and Whitesides~\cite{chaudhury_how_1992} demonstrated that liquid droplets can be driven up a tilted plane against gravity by chemically treating the plane so it gradually becomes more wettable with height.
Later, Ichimura \emph{et al.}~\cite{ichimura_light_2000} showed that sessile droplets start moving in response to gradients in wettability, which are produced by a photo-chemical reaction.
These experiments essentially demonstrated an early use of \emph{structured light} to create motion on the micron scale, an approach which has recently become the focus of intense research~\cite{palagi_structured_2016,varanakkottu_light_2016,grawitter_feedback_2018}.
Light-driven fluid motion~\cite{seki_photo_2005,chevallier_pumping_2011,baigl_photo_2012,schmitt_marangoni_2016,xiao_moving_2018} is an especially favorable control mechanism because of its high precision and controllability through the established experimental methods in optics~\cite{rubinsztein_roadmap_2017}.
Notably, it can be used in combination with or as an alternative to electrowetting techniques~\cite{mugele_electrowetting_2005}.
Precise control of fluid motion is foundational for advanced lab-on-a-chip devices~\cite{squires_microfluidics_2005,darhuber_principles_2005,kaspar_confinement_2016}, as well as self-cleaning surfaces~\cite{blossey_self_2003,edalatpour_managing_2018}, and printing with sub-droplet precision~\cite{li_evaporation_2013,varanakkottu_light_2016}.

Recently, significant advances toward light-switchable substrates with large wettability gradients have been made~\cite{wang_photoresponsive_2007}.
First, Zhu \emph{et al.}~\cite{zhu_light_2020} demonstrated light-induced droplet motion on a substrate modified by azo\-benzene-calix[4]arene where the equilibrium contact angle, as a measure for wettability, ranged from $33^\circ$ to $110^\circ$.
Second, several groups~\cite{lim_photoreversibly_2006,pirani_light_2016} have designed arrays of micropillars made from light-switchable polymer material.
Switching the pillars between an upright and buckled shape offers the possibility to reversibly switch the substrate into a superhydrophic state where wettability is vanishingly small.

In this article we implement the boundary-element method \cite{pozrikidis_boundary_1992,pozrikidis_interfacial_2001,pozrikidis_practical_2002,kim_microhydrodynamics_2005,zinchenko_emulsion_2013,katsikadelis_boundary_2016} to solve the full three-dimensional equations of Stokes flow in order to study droplet motion induced by non-uniform and dynamic wettability patterns.
In particular, we apply our method to moving step profiles in wettability and explore the limits of droplet motion induced by such wettability gradients, \emph{i.e.}, we determine the maximally possible droplet speeds under various conditions.

Literature provides two research lines relevant to us.
First, in two articles McGraw \emph{et al.} implemented the boundary element method to gain insights into the internal flow fields of axisymmetric droplets~\cite{mcgraw_slip_2016,chan_morphological_2017}.
They combined theory and experiment to study how droplets relax towards their new equilibrium shape on a substrate after an instantaneous change in the spatially uniform wettability.
Second, Glasner~\cite{glasner_boundary_2005} implemented the boundary element method to study the dynamics of droplets placed on substrates with static but non-uniform wettability profiles.
The study employed a quasi-static approximation where the gas-liquid interface of the droplet is always assumed to be equilibrated w.r.t.~to the shape of the contact line.
In a very recent example of the same research line, Savva~\emph{et al.}~\cite{savva_droplet_2019} extended Glasner's method by describing the gas-liquid interface with the so-called \emph{thin-film} approximation for Stokes flow for highly wettable substrates.

Our approach goes beyond both research lines because it combines dynamic and spatially non-uniorm wettability patterns and thereby offers the possibility to study continuous droplet motion.
It explores more widely the research avenue recently opened in an experiment by Gao \emph{et al.}~\cite{gao_droplets_2018} who continuously drove a droplet up a tilted plane using a \emph{thermally} induced wettability gradient that moved along with the droplet.

We apply our method to investigate a droplet surfing on a moving step in wettablity in two ways:
First, by positioning the droplet at a constant distance or
offset from the center of the wettability step using a feedback loop, we induce a self-regulated motion of the droplet.
Its velocity depends subtly on the properties of the wettability step and its offset from the droplet.
Second, by letting the wettability step approach the droplet with a constant velocity, we recover a subset of the steady states observed in the previous study with a feedback loop.
This helped us to understand the role of convexity in designing wettability profiles to induce steady droplet motion.

The article is structured as follows.
We present the theory of the boundary element method and its implementation in Sec.~\ref{sec_theory}.
Our numerical approach is validated in Sec.~\ref{sec_validation} by studying a sessile droplet on a substrate with
homogeneous wettability.
We study the droplet surfing on a wettability step using a feedback loop in Sec.\ \ref{sec_feedback} and by moving the wettability step with a constant speed in Sec.~\ref{sec_constant_driving}.
Finally, in Sec.~\ref{sec_conclusions} we present our conclusions.

\section{Boundary element method for dynamic wetting: Theory and implementation}
\label{sec_theory}

\begin{figure}
 \centering
 \includegraphics[width=0.8\linewidth]{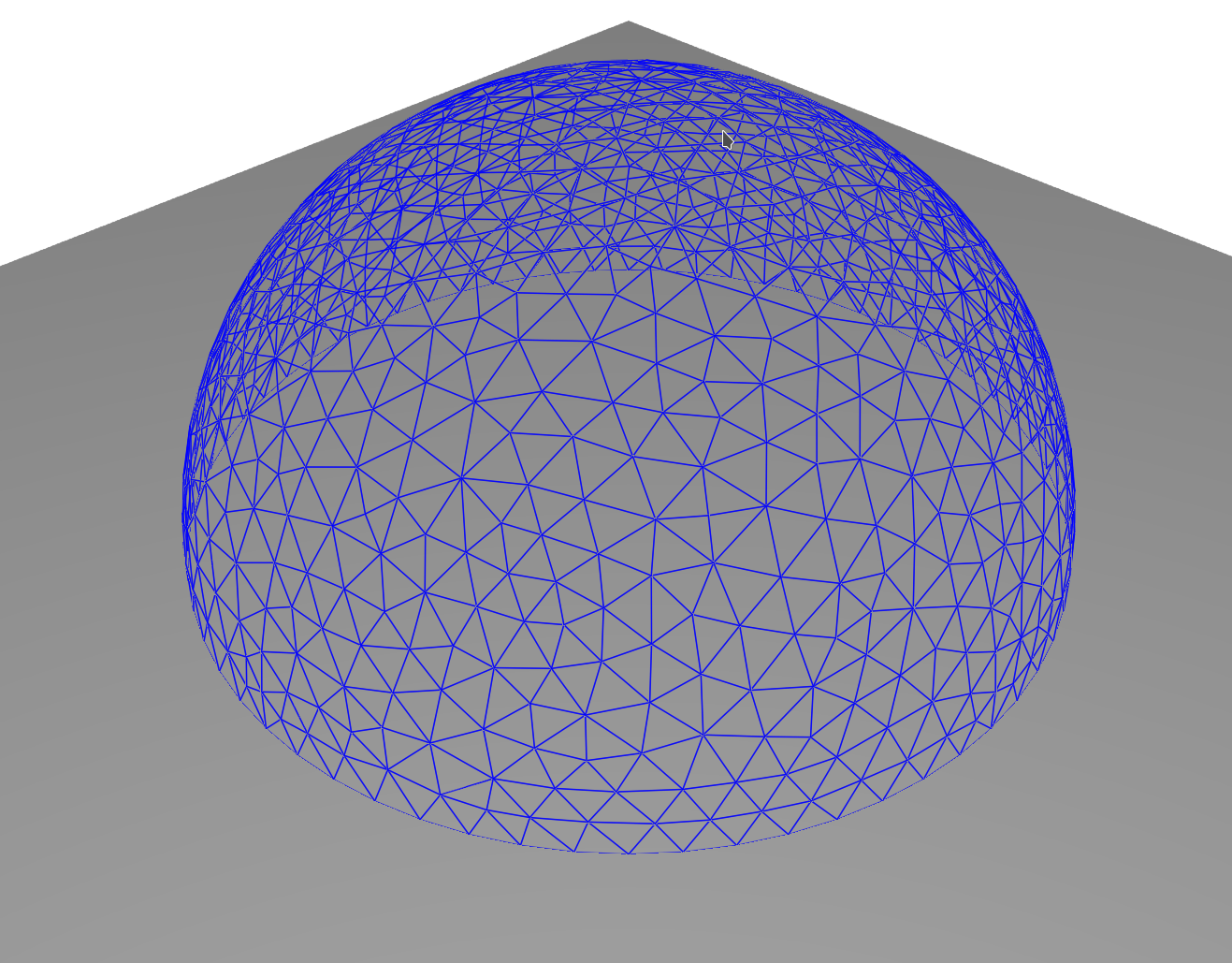}
 \caption{Rendering of a droplet sitting on a substrate (grey surface).
 The blue mesh indicates its \emph{free} surface.}
 \label{fig_sketch}
\end{figure}

We consider a liquid droplet sitting on a solid substrate with switchable wettability (Fig.~\ref{fig_sketch}).
The droplet is bounded by two surfaces:
Its interface with the substrate and its \emph{free} interface with the gaseous atmosphere.
The intersection where the surfaces meet is the contact line, which is the three-phase contact line where liquid, solid, and gaseous phase meet.
Our aim is to describe the droplet motion in response to wettability changes of the substrate.
The current section provides the methodology to treat such a situation.
To move the droplet, we need to calculate the fluid velocity field on the droplet surfaces using the \emph{boundary element method} (Sec.~\ref{subsec.boundary} and the velocity of the contact line using the Cox-Voinov law (Sec.~\ref{subsec.Cox}).
After discretizing the droplet surface (Sec.~\ref{sec_discretization}), we can then move each vertex point~$\vec r^{(i)}$ along the surface normal~$\vec n^{(i)}$ (Sec.~\ref{subsec.time}).
We also implement a method to keep the volume fixed (Sec.~\ref{sec_constraint}) and mesh optimization to avoid acute-angled triangles in the surface mesh (Sec.~\ref{sec_mesh_stab}).
Finally, we nondimensionalize our equations and introduce relevant material parameters (Sec.~\ref{subsec.nondim}).

\subsection{Boundary integral equation for Stokes flow}
\label{subsec.boundary}

The \emph{boundary element method} (BEM) approximates solutions to linear partial differential equations (PDEs)
based on their Green's functions~\cite{pozrikidis_boundary_1992,pozrikidis_practical_2002}.
For droplets moving on substrates we need to study Stokes flow in incompressible fluids, which obeys two linear PDEs for velocity $\vec{v}$ and pressure~$p$,\cite{kim_microhydrodynamics_2005}
\begin{equation}
\mu \nabla^2\vec v(\vec r) = \nabla p(\vec r)
\quad \text{and} \quad
\nabla \cdot \vec v(\vec r) = 0 \, ,
\end{equation}
where $\mu$ is the shear viscosity and $\vec r$ the position vector.
Green's function for Stokes flow in $\mathbb{R}^3$ is the Oseen tensor
\begin{equation}
\tens{O}(\vec{r}) = \frac{1}{8\pi \mu|\vec{r}|} \left( \tens{I} + \frac{\vec{r}\otimes\vec{r}}{\hphantom{^2}|\vec{r}|^2} \right)
\end{equation}
where $\tens I$ is the unit matrix.
Its associated stress field is a tensor of rank three \cite{kim_microhydrodynamics_2005},
\begin{equation}
\tens T(\vec r) = -\frac{3}{4\pi |\vec r|^2} \left(\frac{\vec r \otimes \vec r \otimes \vec r}{\hphantom{^3}|\vec r|^3}\right)
~.
\end{equation}
We are interested in the flow field~$\vec v$ inside and, in particular, on the surface of the closed volume of a droplet.
On any compact region~$D\subset \mathbb{R}^3$ with boundary~$\partial D$ and outward normal described by unit vector~$\vec{n}$, the flow field fulfills the following integral equation~\cite{kim_microhydrodynamics_2005,pozrikidis_boundary_1992,pozrikidis_practical_2002}:
\begin{multline}
f(\vec{r}) \vec{v}(\vec{r}) =
\oint\limits_{\partial D} \tens{O}(\vec{r-r'}) \tens{\sigma}(\vec{r'}) \vec{n}(\vec{r'}) \dif^2\vec{r'}\\
- \oint\limits_{\partial D} \vec{v}(\vec{r'}) \cdot \tens{T}(\vec{r-r'}) \vec{n}(\vec{r'}) \dif^2\vec{r'} \, .
\label{eq_integrals}
\end{multline}
Here, $\tens \sigma$ is the stress tensor of the fluid, $\tens T$ is the stress field of the Oseen tensor introduced above, and $f(\vec r)$ is a dimensionless coefficient, which assumes different values inside~$D$ and at the boundary $\partial D$~\cite{pozrikidis_practical_2002}:
\begin{equation}
f(\vec r) =
\begin{cases}
1 & \text{for } \vec r \in D \setminus \partial D~\text{(inside)}
\\
\frac{1}{2} & \text{for } \vec r \in \partial D \text{, where $\partial D$ is smooth}
\\
\frac{\alpha}{4\pi} & \text{for } \vec r \in \partial D \text{, where $\partial D$ has a corner with}\\&\text{inward solid angle\footnotemark~$\alpha$ \, .}
\end{cases}
\end{equation}%
\footnotetext{The solid angle is calculated locally by considering a small sphere centered at the position $\vec r$ on the droplet surface.
The surface cuts out a volume from the sphere directed towards the droplet, which
encloses the solid angle $\alpha$.
Thus, $\alpha=2\pi$ on smooth parts of the droplet surface and $\alpha=2\theta$ at the contact line, where $\theta$ is the contact angle.}%
The first and the second term on the right-hand side of Eq.~(\ref{eq_integrals}) describe the flow initiated by surface forces $\tens{\sigma}\vec{n}$ and a layer of force/source dipoles at the surface, respectively.
Therefore, they are also called the \emph{single-layer} and \emph{double-layer potentials}.
To close Eq.~(\ref{eq_integrals}), one either needs to eliminate the surface stress force $\tens \sigma \vec n$ or the velocity $\vec v$ by appropriate boundary conditions.
We introduce the boundary conditions in the next paragraph.
Furthermore, in Sec.~\ref{sec_discretization} we describe our discretization of the integral equations so that
approximate values for $\vec{v}$ or $\tens{\sigma} \vec{n}$ can be calculated on each piece of $\partial D$.
Thereafter, we can calculate the velocity field $\vec{v}(\vec{r})$ anywhere in $D$~\cite{pozrikidis_interfacial_2001}.

The Stokes flow problem is only fully determined with boundary conditions.
The droplet boundary consists of two parts, one interface with a solid substrate and a second with a gas phase.
This means each part has its own boundary condition.
At the interface with the substrate there is a Robin boundary condition (for fluids called the Navier condition), $\mu \vec{v} + \lambda\tens{\sigma}\vec{n}= 0$, with a slip length $\lambda$~\cite{bolanos_derivation_2017}, which connects tangential velocity and tangential stress, while the normal component of $\vec{v}$ is zero.
A typical value for the slip length is $\lambda \approx 1\,\text{nm}$~\cite{bonn_wetting_2009} much smaller than the droplet dimension.
At the interface towards the gas phase we assume zero viscosity of the gas phase so that the tangential stress is zero.
There remains a Neumann boundary condition (in hydrostatics the Laplace pressure), $\vec{n} \cdot \tens{\sigma}\vec{n} = - 2 \gamma \kappa$, which locally relates the normal stress component $\vec{n} \cdot \tens{\sigma}\vec{n}$ to the mean curvature $\kappa$, where $\gamma$ is the local surface tension.

\subsection{Cox-Voinov law for the contact line}
\label{subsec.Cox}

Droplets in equilibrium minimize the sum of their surface energies from the interfaces towards the gas phase and substrate, respectively, while preserving their volume~\cite{bonn_wetting_2009}.
If wetting is energetically more favorable than a dry substrate, the liquid phase covers the substrate with a thin film.
Otherwise, a droplet forms with an energetically optimal ratio of the areas between liquid-gas and liquid-substrate interfaces.
On homogeneous and planar substrates, the optimal shape obviously is a spherical cap.
The balance of all forces acting on the three-phase contact line determines the contact angle~$\theta_\mathrm{eq}$ in equilibrium
It results in Young's equation, which relates $\theta_\mathrm{eq}$ to the surface tensions $\gamma_{\mathrm{ij}}$ between liquid (l), gas (g), and substrate (s), respectively:
\begin{equation}
\gamma_\text{sg}=\gamma_\text{sl} + \gamma_\text{lg} \,\text{cos}(\theta_\mathrm{eq}) \enspace.
\label{eq.young}
\end{equation}
Higher wettability means smaller contact angle and according to Eq.~(\ref{eq.young}) that the difference $\gamma_\text{sg} - \gamma_\text{sl}$ of both surface tensions at the substrate increases, since then it is energetically more favorable to cover the substrate with liquid.

Here, when we talk about dynamic wettability patterns, we will vary $\Delta \gamma_\text{s} = \gamma_\text{sg} - \gamma_\text{sl}$ and thereby $\theta_\mathrm{eq}$ in time.
On substrates with heterogeneous wettability the optimal droplet shape can be much more complicated including complex shapes of the contact line and varying curvature of the liquid-gas interface~\cite{glasner_boundary_2005}.

The dynamics of the contact line and the contact angle are not directly determined within the approach outlined so far.
Because the droplet surface is not smooth at the contact line, it is not clear which boundary condition to use here.
For the Neumann boundary condition towards the gas phase the mean curvature $\kappa$ diverges, while the Navier condition at the substrate does not contain
any surface tension necessary to obtain Young's equation (\ref{eq.young}) in equilibrium.
Therefore, the contact line dynamics has to be determined by an additional relation, which translates the fluid motion in the microscopic contact line region to the macroscopic scale.
Several such relations for dynamic wetting exist~\cite{bonn_wetting_2009,eral_contact_2013,ledesma_theory_2013,snoeijer_moving_2013,voinov_hydrodynamics_1976,cox_dynamics_1986}.
The most well-known among them is the Cox-Voinov law~\cite{voinov_hydrodynamics_1976,cox_dynamics_1986} derived from hydrodynamic considerations:
\begin{equation}
v_\text{contact} = \frac{\gamma_\text{lg}}{9\mu\ln(h/\lambda)} \left( \theta_\text{dyn}^3 -\theta_\mathrm{eq}^3 \right) \, .
\label{eq.cox}
\end{equation}
It relates the difference of the cubes of dynamic and equilibrium contact angles, $\theta_\text{dyn}$ and $\theta_\mathrm{eq}$, to the velocity of the contact line $v_\text{contact}$.
The dimensionless coefficient $\ln(h/\lambda)$ is given by the ratio of slip length~$\lambda$ to a macroscopic length scale~$h$.
Voinov defines $h$ as the height above the substrate at which $\theta_\mathrm{dyn}$ is measured.
However, experiments have demonstrated that $\ln(h/\lambda)$ should be treated as a free parameter characterizing the mobility of the contact line~\cite{deruijter_contact_1997}.
The velocity $v_\text{contact}$ determines the component of the
fluid velocity perpendicular to the contact line and in the plane of the substrate.
We evaluate $\theta_\text{dyn}$ using the surface normal $\vec n$ and the substrate normal $\vec e_z$ at the contact line:
\begin{equation}
\text{cos}\, \theta_\text{dyn}  = \vec{n}(t) \cdot \vec e_z \, .
\end{equation}
The resulting velocity $v_\text{contact}$ of Eq.~(\ref{eq.cox}) is included in the discretized BEM problem as a boundary condition by introducing a corresponding stress at the contact line as an additional unknown.
Furthermore, in the direction of the substrate normal we prescribe a vanishing velocity and along the tangent of the contact line we set the stress to zero.
Together, these three conditions fully determine fluid velocity and stress along the contact line.%
\footnote{A similar approach was previously applied to bubbles on a submerged substrate~\cite{pityuk_boundary_2018} although the dynamics of submerged contact lines are necessarily distinct from our case.}

On a light-switchable substrate the light patterns affect the surface tension difference $\Delta \gamma_\text{s} = \gamma_\text{sg} - \gamma_\text{sl}$ in Eq.~(\ref{eq.young}).
Therefore, as mentioned already, we will treat $\varDelta\gamma_\text{s}$ as a heterogeneous and dynamic quantity, which locally determines $\theta_\mathrm{eq}$ and, thereby, deforms the droplet.

\subsection{Discretization}
\label{sec_discretization}
\begin{figure}
\centering
 \includegraphics[width=0.6\linewidth]{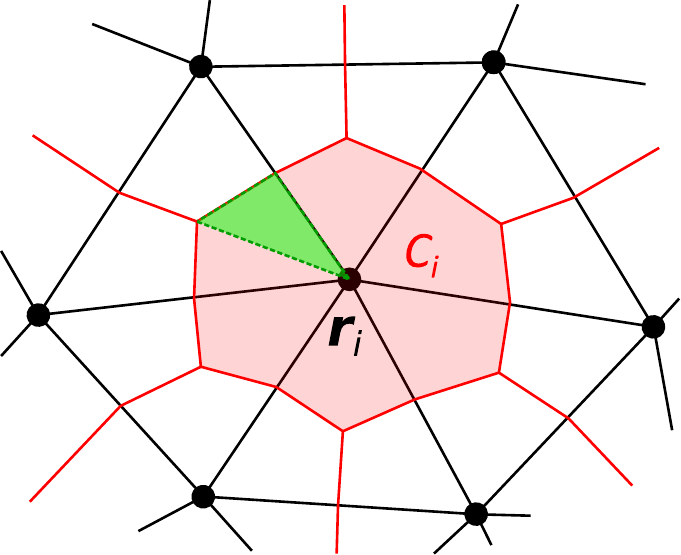}
 \caption{Schematic of triangular mesh (black) in the neighborhood of vertex~$\vec r^{(i)}$ and its associated polygonal region~$C_i$ (red).
The region $C_i$ is defined by and subdivided into triangles, one of which is indicated in green.
The three corners of the triangles are the central vertex~$\vec r^{(i)}$, the point halfway toward a neighboring
vertex~$\vec r^{(j)}$, and the centroid (barycenter) of one of the black mesh triangles.
}
 \label{fig_mesh}
\end{figure}
We discretize Eq.~(\ref{eq_integrals}) using a triangular mesh to cover all droplet surfaces (solid-liquid and vapour-liquid).
From the triangular mesh we define a dual mesh in which each vertex at position
$\vec r^{(i)}$ is surrounded by a compact polygonal region~$C_i=C(\vec r^{(i)})$ (Fig.~\ref{fig_mesh}).
Together, all the regions $C_i$ cover the entire surface.
For a sufficiently fine and well-adapted mesh, field variables are approximately constant on each $C_i$ and integrals can thus be evaluated piece-wisely.
As a result, Eq.~(\ref{eq_integrals}) becomes a system of linear equations of the form:
\begin{equation}
 \sum_j \tens A^{(ij)} \vec v^{(j)} = \sum_j \tens B^{(ij)} \vec t^{(j)} \, ,
 \label{eq_linear}
\end{equation}
where $\vec v^{(j)}$ and $\vec t^{(j)} = \tens \sigma^{(j)} \vec n^{(j)}$ are the  respective values of the flow field and the stress or surface force at vertex $j$.
The tensors $\tens A^{(ij)}$ and $\tens B^{(ij)}$ are piece-wise integrals, i.e.,
\begin{equation}
\tens A^{(ij)} =  f(\vec r_i)\delta_{ij} \,\tens I + \int\limits_{C_j} \tens{T}(\vec{r}^{(i)}-\vec r') \vec{n}(\vec{r'}) \dif^2\vec{r'}
\label{eq_vel_bem}
\end{equation}
with Kronecker symbol $\delta_{ij}$ and identity matrix $\tens I$, and
\begin{equation}
\tens B^{(ij)} =  -\int\limits_{C_j} \tens{O}(\vec{r}^{(i)}-\vec r') \dif^2\vec{r'}~.
\label{eq_stress_bem}
\end{equation}
The linear algebra problem in Eq.~(\ref{eq_linear}) is fully solvable by eliminating either $\vec v^{(j)}$ or $\vec t^{(j)}$ with the help of boundary conditions at each vertex.
It will be convenient to collect all matrices $\tens A^{(ij)}$ and $\tens B^{(ij)}$ into block matrices $\tens A$ and $\tens B$ and all vertex velocities and stress vectors into combined vectors $\vec v$ and $\vec t$, so that Eq.~(\ref{eq_linear}) can be written simply as
\begin{equation}
 \tens A \vec v = \tens B \vec t~.
 \label{eq_linear_simple}
\end{equation}

To perform the integrations in Eqs.~(\ref{eq_vel_bem}) and (\ref{eq_stress_bem}), we subdivide the polygonal region $C_i$ into triangles (Fig.~\ref{fig_mesh}) and numerically integrate over each triangle  using a Gaussian quadrature stencil with $9$~points for nonsingular integrands and with $400$~points for singular integrands (i.e.~when $i=j$ in Eqs.~(\ref{eq_vel_bem}) and (\ref{eq_stress_bem}))~\cite{pozrikidis_boundary_1992,katsikadelis_boundary_2016}.
Note that we never evaluate the integrand at the vertex $\vec r^{(i)}$, where the tensors $\tens O$ or $\tens T$ have integrable singularities, but always choose points inside the triangles for performing the Gaussian quadrature, which is sufficient.

To solve Eq.~(\ref{eq_linear_simple}), we first apply the boundary conditions introduced above to remove one of the variables at the vertex, either velocity or surface force.
Specifically, at the interface towards the gas phase the tangential component of the surface force is zero while we replace the normal component by the Laplace pressure~$-2 \gamma \kappa$.
To calculate the principal curvatures at each vertex, we use the discrete Laplace-Beltrami operator described in Method~C of the review of Guckenberger \emph{et al.}~\cite{guckenberger_bending_2016} and the normal vector~$\vec n^{(i)}$ follows by taking the sum of the normal vectors of all neighboring triangles and normalizing it.
At the interface with the solid substrate we set the normal velocity component to zero and eliminate the tangential component of the surface force by the Robin boundary condition.
Finally, at the contact line the velocity component normal to the substrate is zero.
The in-plane velocity component normal to the contact line obtains the value prescribed by the Cox-Voinov law, and the tangential component of the surface force is zero.
We then collect all the remaining unknown variables on the left-hand side and all the known quantites on the right-hand side and solve the resulting system of linear equations by inversion of the coefficient matrix.
Finally, the missing velocity and surface-force variables are calculated from the boundary conditions.
In total, solving the linear problem of Eq.~(\ref{eq_linear_simple}) yields approximations for $\vec v$ and $\tens  \sigma  \vec n$ anywhere on the droplet surface.
Once these quantities are known, $\vec v(\vec r)$ can be calculated anywhere within the droplet by numerical integration from Eq.~(\ref{eq_integrals}).

\subsection{Time evolution}
\label{subsec.time}

In the discretized version the dynamics of the free droplet surface is specified by the velocity of each vertex with position vector~$\vec r_i$.
Concretely, the vertex moves with the local normal component of the fluid velocity and an artificial tangential component~$\vec w^{(i)}_\mathrm{t}$, which is due to mesh optimization and introduced rigorously in Sec.~\ref{sec_mesh_stab},
\begin{equation}
\hphantom{\quad .}
\frac{\mathrm{d} \vec r^{(i)}}{\mathrm{d} t}
 = (\vec n^{(i)} \otimes \vec n^{(i)}) \vec v^{(i)}
 + \vec w^{(i)}_\mathrm{t}
\, .
\label{eq_surface_dynamics}
\end{equation}
Motion tangential to the droplet's free surface is physically irrelevant because it does not affect the shape of the droplet.
However, tangential motion can be used to
improve \emph{mesh fitness}, i.e., adaptation of the mesh to the droplet shape.

Note that time appears only through Eq.~(\ref{eq_surface_dynamics}) in our problem because Stokes flow adapts instantaneously to the droplet's shape.
Numerically, the dynamics of each $\vec{r}^{(i)}$ is calculated using a state-of-the-art Runge-Kutta algorithm~\cite{tsitouras_runge_2011} once the velocity vectors $ \vec v^{(i)}$ are calculated with the BEM as described above.

We refine the numerical setting described so far with two additional techniques described in Secs.~\ref{sec_constraint} and \ref{sec_mesh_stab}: a volume constraint and a mesh optimization method.

\subsection{Volume constraint}
\label{sec_constraint}
We follow a method first suggested by Alinovi \emph{et al.} \cite{alinovi_boundary_2017} to preserve the volume of the droplet during motion through a side condition for the velocity vectors of the droplet surface.
Since volume conservation can be formulated by the chain rule, we obtain
\begin{equation}
    \frac{\mathrm{d}V}{\mathrm{d}t} =
    \sum\limits_i \nabla_{\vec r^{(i)}} V  \cdot \frac{\mathrm{d} \vec r_i}{\mathrm{d} t}
    = \sum\limits_i \nabla_{\vec r^{(i)}} V \cdot \vec v^{(i)} = 0~,
\label{eq.constraint}
\end{equation}
We call $\vec c^{(i)}= \nabla_{\vec r^{(i)}}
V$ the local constraint vector and for further use we concatenate all $\vec c^{(i)}$ into a vector $\vec c$, in the same order as $\vec v$ and $\vec t$ in Sec.~\ref{sec_discretization}).
Alinovi \emph{et al.} estimate $\vec c^{(i)}$ as the product of the local normal $\vec n^{(i)}$ vector and the surface element, which requires a smooth surface around each vertex.
\footnote{This estimate makes sense since $\nabla_{\vec r^{(i)}} V$ is the direction of maximal change in $V$  and since any volume change must be proportional to the area associated with a given vertex.}
In order to also treat the kink in the droplet surface at the contact line, we instead calculate the gradients
$\nabla_{\vec r^{(i)}} V$ directly by numerical differentiation, which gives well-defined constraints for the contact line.
Now, with the introduction of a Lagrange multiplier~$\varLambda$ for the constraint of Eq.~(\ref{eq.constraint}), Eq.~(\ref{eq_linear_simple}) is extended to
\begin{equation}
    \begin{pmatrix}
    \tens A & \vec c \\
    \vec c^\mathrm{T} & 0
    \end{pmatrix}
    \begin{pmatrix}
    \vec v \\
    \varLambda
    \end{pmatrix}
    =
    \begin{pmatrix}
    \tens B \vec t\\
    0
    \end{pmatrix}~.
\end{equation}
Any solution for the flow field on the droplet surface is now guaranteed to conserve volume for an infinitesimal step in direction of $\vec v$.

\subsection{Mesh optimization}
\label{sec_mesh_stab}

During the simulation we optimize the mesh representing  the droplet meaning we avoid acute-angled triangles and keep the triangle areas uniform across the whole droplet surface.
This guarantees that discretization errors remain minimal.
According to Zinchenko \emph{et al.} \cite{zinchenko_emulsion_2013}, a mesh fitness is introduced that is maximized for equilateral triangles.
It is a function of all the edge lengths of the triangles and their areas.
Thus, improving the mesh fitness  w.r.t.~the triangle shape minimizes discretization errors.

Following Zinchenko \emph{et al.} \cite{zinchenko_emulsion_2013}, we apply two measures at different phases of time integration.
First, when calculating the r.h.s.~of Eq.~(\ref{eq_surface_dynamics}), we choose the tangential velocity of each vertex on the free droplet surface including vertices on the contact line such that the rate of change of the overall mesh fitness is minimal.
\footnote{Zinchenko \emph{et al.}~call this approach \emph{passive mesh stabilization}.}
We then move all the vertices on the free droplet surface by one Runge-Kutta time step.
Note that the normal velocity, which affects droplet shape and motion, remains unchanged.
Second, as a next step we check the fitness of the mesh facing the substrate and if the maximum edge lengths exceeds $150\,\%$ of the minimal edge length,  the positions of all substrate vertices (without the ones on the contact line) are adjusted by tangential displacements.
\footnote{Zinchenko \emph{et al.} call this approach \emph{active mesh stabilization}.}

Since this technique is completely derived from literature and by its very nature must not affect the dynamics  of the droplet, we do not reproduce the algorithm in detail here.
In Appendix~\ref{sec_mesh_benchmark}, Fig.~\ref{fig_mesh_fitness_glycerin} we
display mesh fitness over time for a benchmark case to show the effectiveness of
the implemented method.

\subsection{Nondimensionalization and material parameters}
\label{subsec.nondim}

\begin{table*}
\centering
 \begin{tabular}{c|c|r|r|r|r|r|r|r|r|r|r|r}
\# & material & $\mu$ & $\nu$ & $\gamma$ & $\lambda$ & $R_0$ & $\tau$ & $f_c$ & $\tilde \gamma$ & $\tilde \lambda$ & $\ln(h/\lambda)$ & $\tau_c / \tau$\\
   &  & [$\mathrm{mPa}\!\cdot\!\mathrm{s}$] & [$\frac{\mathrm{mm}^2}{\mathrm{s}}$] & [$\frac{\mathrm{mN}}{\mathrm{m}}$] & [$\mathrm{nm}$] & [$\mu\mathrm{m}$] & [$\mu\mathrm{s}$] & [$\mu\mathrm{N}$] &  &  &  & \\
  \hline
  1 & $90\,\%$ glycerol & $209$ & $169$ & $65.3$ & $1$ & $100$ & $59$ & $35$ & $0.19$ & $10^{-5}$ & 44$^\ast$ & $2084$\\
  2 & $90\,\%$ glycerol & $209$ & $169$ & $65.3$ & $1$ & $1500$ & $13,300$ & $35$ & $2.8$ & $0.67\cdot 10^{-6}$ & 44$^\ast$ & $141$\\
 \end{tabular}
\caption{Material parameters: dynamic viscosity~$\mu$, kinematic viscosity~$\nu$, surface tension~$\gamma$, slip length~$\lambda$, unit of length~$R_0$, unit of time~$\tau$, unit of force~$f_c$, dimensionless surface tension~$\tilde \gamma$, dimensionless slip length~$\tilde \lambda$, Cox-Voinov-coefficient~$\ln(h/\lambda)$, and characteristic time~$\tau_c$ for $\theta=90^\circ$.
$^\ast$For glycerin, de Ruijter \emph{et al.}\cite{deruijter_contact_1997} fitted $\theta_\mathrm{dyn}(t)$ as a solution of Eq.~(\ref{eq_spherical}) to their experimental data and observed the value $44$ for $\ln(h/\lambda)$.
They interpret this large value as the result of ``an additional source of energy dissipation within the three-phase zone''\cite{deruijter_contact_1997}.%
}
\label{tab_parameters}
\end{table*}

We are able to nondimensionalize our continuum equations by introducing a characteristic length scale~$R_0$, a time scale $\tau$, and a force scale $f_c$.
Using these three parameters, we will then rewrite all quantities $a_i$ as dimensionless quantities~$\tilde a_i$.
For $R_0$ we choose the radius of the initial circular base area of the droplet.
Furthermore, we remove dynamic and kinematic viscosities from our equations, which then implies the characteristic time and force scales
\begin{equation}
 \tau = \frac{R_0^2}{\nu} \quad \text{and}  \quad f_c = \nu\mu~.
\end{equation}
Note that $f_c$ is the intrinsic force scale of a Newtonian fluid and the limit of small Reynolds number means that all acting forces are smaller than $f_c$.
All relevant quantities can now be written in dimensionless form: curvature~$\tilde \kappa = R_0\kappa$, slip length~$\tilde \lambda = \lambda / R_0$, stress tensor components $\tilde{\sigma}_{ij} = \sigma_{ij} R_0^2 / f_c$, and surface tensions $\tilde \gamma_{\text{ij}}=  \gamma_{\text{ij}} R_0 / f_c$.
Since the reduced time and coordinates are $\tilde t = t/\tau$ and $\tilde x = x / R_0$, the respective reduced derivatives become $\tilde \partial_t = \tau \partial_t$ and $\tilde \partial_x = R_0 \partial_x$.

In Ref.~\cite{deruijter_contact_1997} Ruijter \emph{et al.} considered the relaxation of a droplet on a uniform substrate towards its equilibrium shape assuming that the droplet is always a spherical cap (spherical cap model).
Then, the constant volume $V$ or its reduced value $\tilde V$ is always strictly related to the momentary contact angle.
In particular, with the initial contact angle $\theta_0$ we have
\begin{equation}
\tilde V = \frac{V}{R_0^3} =\frac{\pi}{3} \frac{(2 + \cos \theta_0)(1 - \cos \theta_0)^2}{\sin^3 \theta_0} \, .
\end{equation}
For the half-sphere droplet with $\theta_0=\pi/2$, this gives $\tilde V = 2\pi/3$, as it should.
Assuming that the dynamics of the relaxing droplet is completely determined by the motion of the contact line, the authors identify a characteristic relaxation time $\tau_c$, which in units of $\tau$ reads
\begin{equation}
\tau_c / \tau = -9 \, \tilde{\gamma}_\text{lg}^{-1} \, \sqrt[3]{3 \tilde V/\pi} \, \ln\Big(\tilde \lambda /
\sqrt[3]{3\tilde V / 2\pi}\Big) \, .
\label{eq.tau_c}
\end{equation}

In Table~\ref{tab_parameters} we collect the material parameters and their dimensionless counterparts for a 90\,\%~glycerine/10\,\%~water mixture for various length values $R_0$ and use them in the following.

\section{Validation: Switching a homogeneous substrate}
\label{sec_validation}

\begin{figure}
 \flushright
 \begin{overpic}[width=0.96\linewidth]{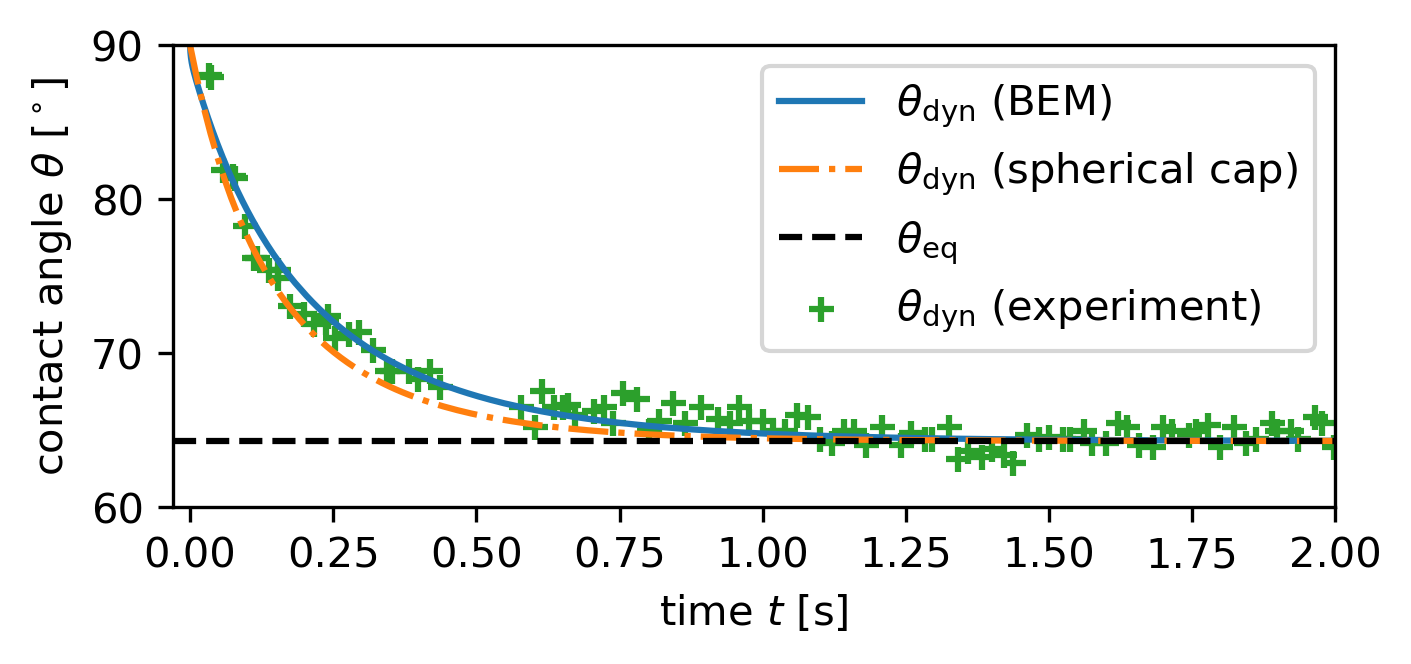}
     \put(-5,42){\Large\textsf{(a)}}
 \end{overpic}
  \begin{overpic}[width=\linewidth]{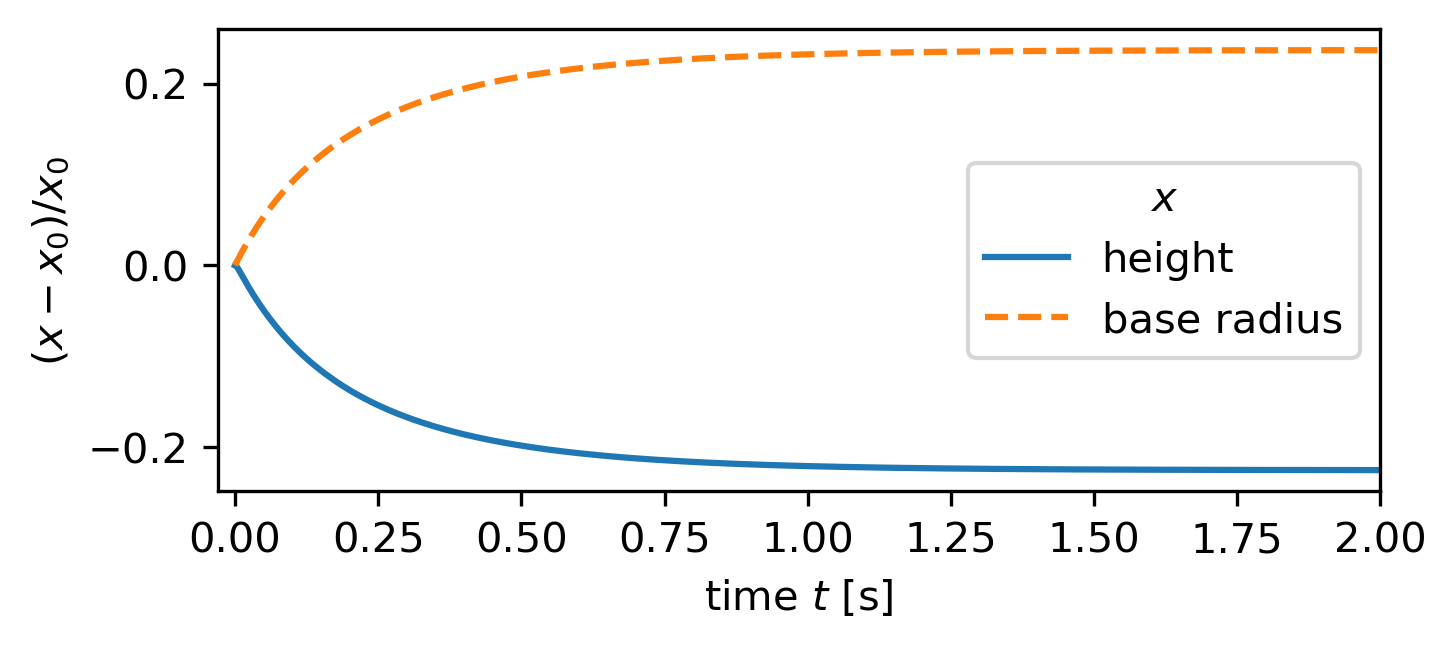}
     \put(0,40){\Large\textsf{(b)}}
  \end{overpic}
 \caption{
Relaxation of a 90\,\%~glycerol droplet after the wettability of the substrate is switched from an equilibrium contact angle of $90^{\circ}$ to $\theta_\mathrm{eq}=65^\circ$ at time $t=0$:
a) Dynamics of the contact angle $\theta_\mathrm{dyn}$ plotted versus time.
Solid blue line shows the result from the BEM and dashed-dotted orange line from the spherical cap model.
The dashed black line shows the final equilibrium contact angle.
The symbols refer to experimental data points reproduced from Fig.~2(a) in Ref.~\cite{deruijter_contact_1997}.
b) Deformation of the droplet as a function of time either characterized by droplet height (solid blue line) or radius of the base area (dashed orange line).
The change relative to the respective initial values $x_0$ are shown.
Both curves are calculated with the BEM.
}
 \label{fig_glycerin}
\end{figure}

We apply our method to a simple case to validate its performance.
The simplest applicable test case is the relaxation of a sessile droplet on a homogeneous substrate that switches
from one wettability to another (see movie~M01 in the ESI$^\dag$).
Thus the equilibrium contact angle is switched at a specific time and we observe how the droplet relaxes towards the
new equilibrium contact angle.
Note that our method includes a valid model for the motion of the contact line because we impose the Cox-Voinov law explicitly.
However, the motion of the contact line is coupled to the motion of the free
surface and the fluid inside the droplet.

In the following we illustrate the overall dynamics of the droplet at a specific system.
We rely on experimental data published by de Ruijter \emph{et al.}\cite{deruijter_contact_1997} who observed the spreading of ca.~$5$-$10\,\mathrm{mm}^3$ of a $90\,\%$ glycerol/$10\,\%$ water mixture on PET.
All material properties and parameters are collected in Table~\ref{tab_parameters}, row~2.

\begin{figure}
\centering
\includegraphics[width=\linewidth]{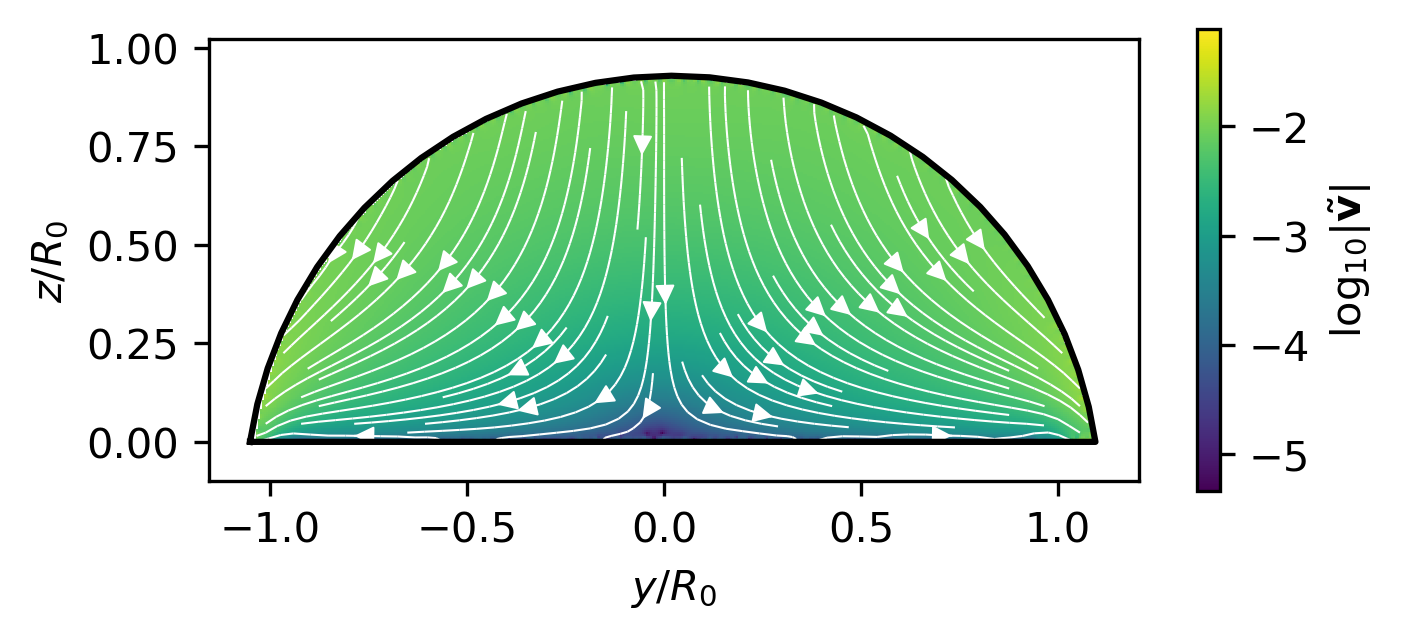}
\caption{%
Computed stream lines (white solid) in the central cross section of the  $90\,\%$ glycerol droplet at $t=7\,\tau$.
Its outline is the solid black line and background color indicates speed.
The streamlines follow within the BEM.
}
 \label{fig_glycerin_streamlines}
\end{figure}

As already explained in Sec.~\ref{subsec.nondim} De Ruijter \emph{et al.} modeled the relaxation of the droplet using the spherical-cap model and used the Cox-Voinov law to completely describe the droplet dynamics by the dynamic contact angle~$\theta_\mathrm{dyn}$.
It obeys the following differential equation:
\begin{multline}
 \frac{\mathrm{d}\theta_\mathrm{dyn}}{\mathrm{d}t} =
 -\frac{\gamma_\text{lg}}{9\mu \ln(h/\lambda)}\sqrt[3]{\frac{\pi}{3V}}
 (\theta_\mathrm{dyn}^3 - \theta_\mathrm{eq}^3) \\
\times \sqrt[3]{(1-\cos\theta_\mathrm{dyn})^2\;(2+\cos\theta_\mathrm{dyn})^4}
 \label{eq_spherical}
\end{multline}
The characteristic time $\tau_c$ introduced in Eq.~(\ref{eq.tau_c}) follows by linearizing Eq.~(\ref{eq_spherical}) in small changes of $\theta_\mathrm{dyn}$ and including only the coefficients independent of $\theta_\mathrm{dyn}$ or $\theta_\mathrm{eq}$.

In Fig.~\ref{fig_glycerin} we present results for the relaxing glycerin droplet using the material parameters of row~2 in Table~\ref{tab_parameters}.
We compare our results for the contact angle from the BEM to numerical solutions of Eq.~(\ref{eq_spherical}) in Fig.~\ref{fig_glycerin}(a).
We find qualitative agreement between both approaches.
However, the observable deviations of the two curves, the BEM shows a slower relaxation in the contact angle, reveals that the fluid flow included in our method slows the relaxation further down.
Our simulated dynamics also matches the experimental results in both the shape of the relaxation curve of the contact angle and relaxation time, as the experimental points in Fig.~\ref{fig_glycerin}(a) show.
We achieve this despite several uncertainties:
We know neither the exact initial shape of the droplet in the experiments of de Ruijter \emph{et al.} nor the exact volume of liquid they used.
We also note during our simulations fluctuations in the droplet volume do not exceed $0.15\,\%$ of the initial volume while the droplet base radius and height, displayed in Fig.~\ref{fig_glycerin}(b), change by over $20\,\%$ of their initial values.
This cearly demonstrates the robustness of our implementation.
Finally, in Fig.~\ref{fig_glycerin_streamlines} we display the stream lines of the fluid at $t=7\tau$, along which material is transported from the top to the sides of the droplet as it spreads.

\begin{figure}
 \includegraphics[width=\linewidth]{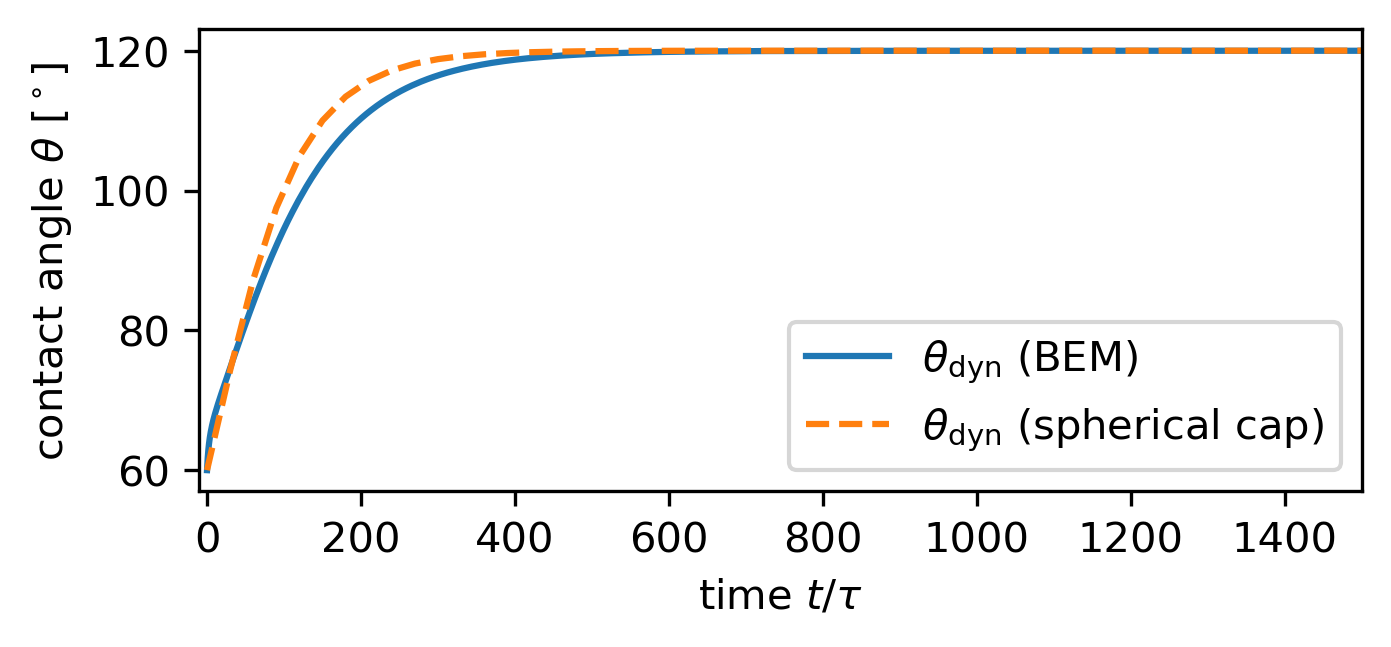}
 \caption{%
Relaxation of the contact angle from $60^\circ$ to $120^\circ$ of a $90\,\%$ glycerol/$10\,\%$ water droplet with $R_0=100\mu\mathrm{m}$ on a homogeneous substrate.%
 }
 \label{fig_60_120_relaxation}
\end{figure}

We also studied a number of wettability switches on homogeneous substrates to test the technical limitations of our mesh stabilization.
Specifically, we were able to realize changes in contact angle of $\pm15^\circ$ when starting from $60^\circ$ and $\pm30^\circ$ when starting from $90^\circ$.
Notably, the latter case means that our method is able to cross into regions where droplets overhang their base area without any special modifications to treat this case.
For example, in Fig.~(\ref{fig_60_120_relaxation}) we demonstrate the relaxation of the contact angle from $\theta_\text{dyn} = 60^\circ$ to  $\theta_\text{eq} = 120^\circ$.
(A corresponding animation is included as movie~M01 in the ESI.$^\dag$)
A clear deviation from the spherical-cap model is visible.

\section{Surfing on a wettability step profile with feedback}
\label{sec_feedback}

\subsection{Setup}
\begin{figure}
\centering
  \includegraphics[width=0.95\linewidth]{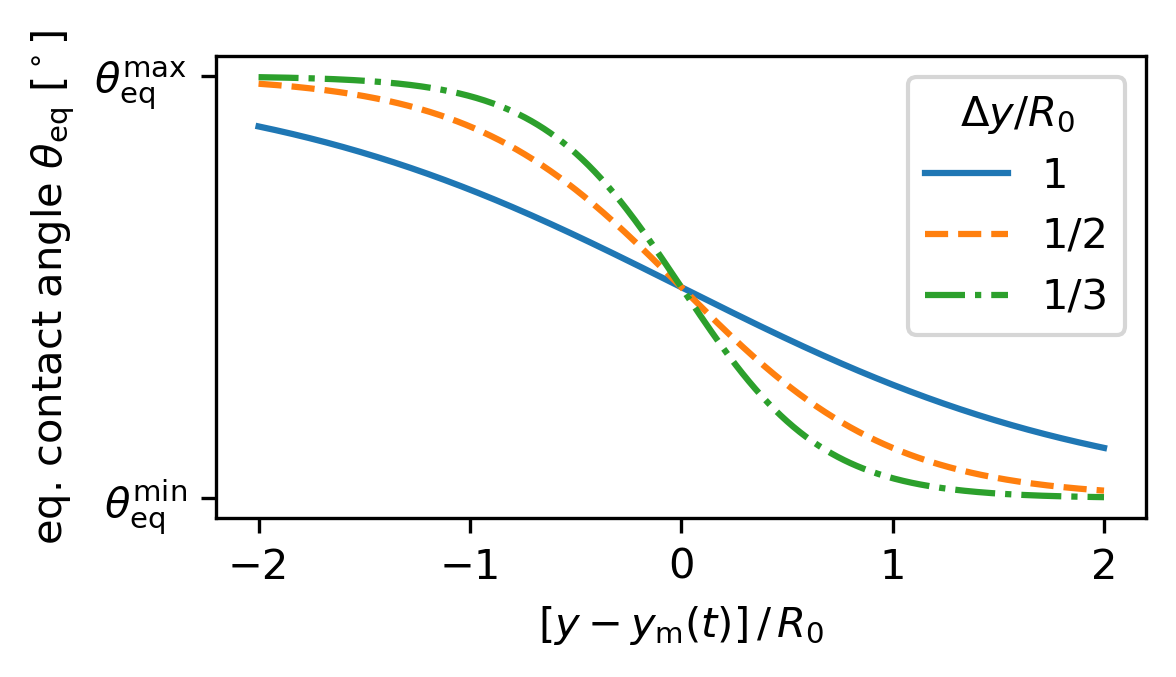}
  \caption{Three examples for a step profile in wettability quantified according to Eq.~(\ref{eq_gradient}) by the equilibrium contact angle.
  $y_m$ is the center of the wettability step with maximal gradient and $\Delta y$ its width.
}
  \label{fig_feedback_schematic}
\end{figure}

A droplet placed on a substrate with a gradient in wettability will move in the direction of increasing wettability meaning smaller contact angle.
In the following we consider a smooth step profile in wettability (\emph{cf.} Fig.~\ref{fig_feedback_schematic}) and
move it with the droplet so that the distance between wettability step and the droplet's center of mass stays constant.
Thus, the droplet quasi ``surfes'' on the wettability step (see movie~M02 in ESI$^\dag$).
Using such a feedback between droplet motion and wettability step, we will determine the maximum speed achievable by droplets moving in a gradient of fixed steepness.

In our approach we quantify the wettability of the substrate by the local equilibrium contact angle.
To realize a step profile in wettability,  $\theta_\mathrm{eq}(y, t)$ (displayed in Fig.~\ref{fig_feedback_schematic}), we use the logistic step function and write
\begin{equation}
 \theta_\mathrm{eq}(y, t) = \theta_\mathrm{eq}^\mathrm{max}
-
\frac{\theta_\mathrm{eq}^\mathrm{max} - \theta_\mathrm{eq}^\mathrm{min}}{1 + \exp(-[y - y_\mathrm{m}(t) ] / \varDelta y)}
  \label{eq_gradient}
\end{equation}
Thus, the wettability step is characterized by its maxmimum and mimimum contact angles~$\theta_\mathrm{eq}^\mathrm{max}$ and $\theta_\mathrm{eq}^\mathrm{min}$ and by the width $\varDelta y$ as a measure for its maximum steepness.
Via feedback control we keep the position of the step center at $y_m(t) = y_c(t) -s$, where $y_c(t)$ is the position of the droplet center of mass and $s$ the fixed offset between the step and droplet centers.
Here,
$s < 0$ means that the droplet
lags behind the wettability step.
In the following, we study how the four parameters $\theta_\mathrm{eq}^\mathrm{max}$, $\theta_\mathrm{eq}^\mathrm{min}$, $\varDelta y$, and $s$ influence the droplet speed.
In addition, we analyse the deformation of the droplet to a nonequilibrium but steady shape while in motion and the steady-state flow field inside the droplet.

Here, all droplets are simulated using the material properties set out in row~1 of Table~\ref{tab_parameters} which describes a $90\,\%$ glycerol/$10\,\%$ water mixture with $R_0=100\,\mu\mathrm{m}$.

\begin{figure*}
  \parbox[b]{0.48\linewidth}{
  \centering
  \begin{overpic}[width=0.35\linewidth]{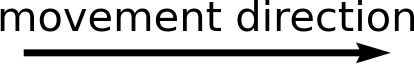}
   \put(-100,0){\Large\textsf{(a)}}
  \end{overpic}
  \flushright
  \includegraphics[width=\linewidth]{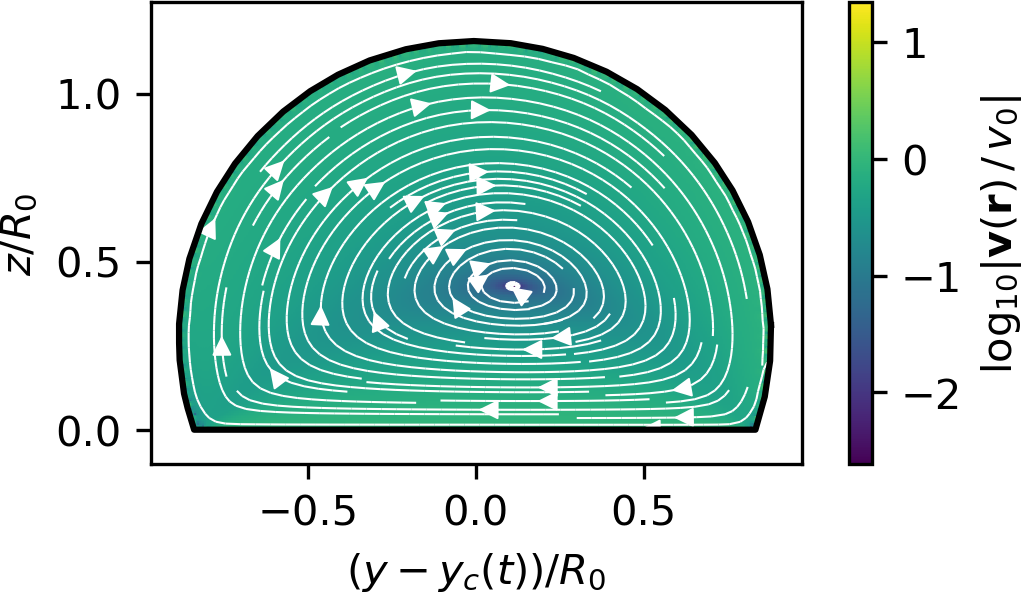}
  }\quad%
\parbox[b]{0.5\linewidth}{
  \flushright
  \phantom{\includegraphics[width=0.35\linewidth]{feedback_direction}}
  \begin{overpic}[width=0.95\linewidth]{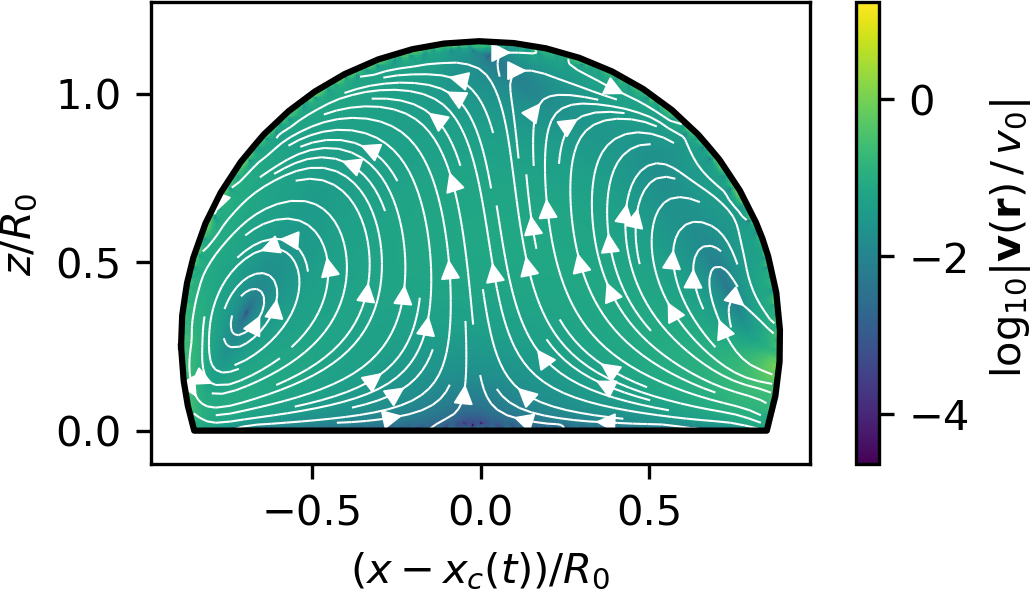}
   \put(0,61){\Large\textsf{(b)}}
  \end{overpic}
  }
 \caption{%
Flow field in the interior of the droplet (a) in the comoving reference frame of the droplet moving with speed~$v_0$, and (b) in a cross-sectional plane perpendicular to the direction of motion for a $90\,\%$ glycerol/$10\,\%$ water droplet with $R_0=100\,\mu\mathrm{m}$ ``surfing'' with $\theta_\mathrm{eq}^\mathrm{min}=90^\circ$, $\theta_\mathrm{eq}^\mathrm{max}=120^\circ$, $\varDelta y = 1/3$, and $s=0$.
An animation of point-like tracer particles corresponding to (a) is contained in the ESI$^\dag$ as movie~M05.
}
 \label{fig_feedback_flow}
\end{figure*}

\subsection{Flow field}

Regardless of the specific values for each parameter for the wettability step, the droplet eventually settles into a steady state with constant shape and constant flow field.
The moving droplet never adjusts to the local equilibrium angles exactly.
Rather a difference remains between $\theta_\mathrm{dyn}$ and $\theta_\mathrm{eq}$, which drives the droplet and determines its steady-state speed
\begin{equation}
   v_0
  =  \lim\limits_{t\to\infty}\frac{\mathrm{d} y_\mathrm{c}}{\mathrm{d}t\;}
\quad.
 \end{equation}
Notably, the droplets never enter into a limit cycle with oscillating speed but rather approach a speed value constant in time.

As a representative example, in Fig.~\ref{fig_feedback_flow} we display the internal flow field of a droplet surfing on a wettability step characterized by $\theta_\mathrm{eq}^\mathrm{min} = 90^\circ$, $\theta_\mathrm{eq}^\mathrm{max} = 120^\circ$, and $\varDelta y = 1/3$ at an offset~$s=0$, \emph{i.e.}, directly on the steepest location.
The flow field forms one single vortex filling the droplet in the plane parallel to its direction of motion and two vortices in the plane perpendicular to this direction.
The fluid is transported predominantly along the free surface and so, in effect, it \emph{rolls} on the substrate as the droplet moves forward.

\subsection{Speed}
\label{sec_feedback_speed}
\begin{figure}
  \flushright
 \includegraphics[width=0.93\linewidth]{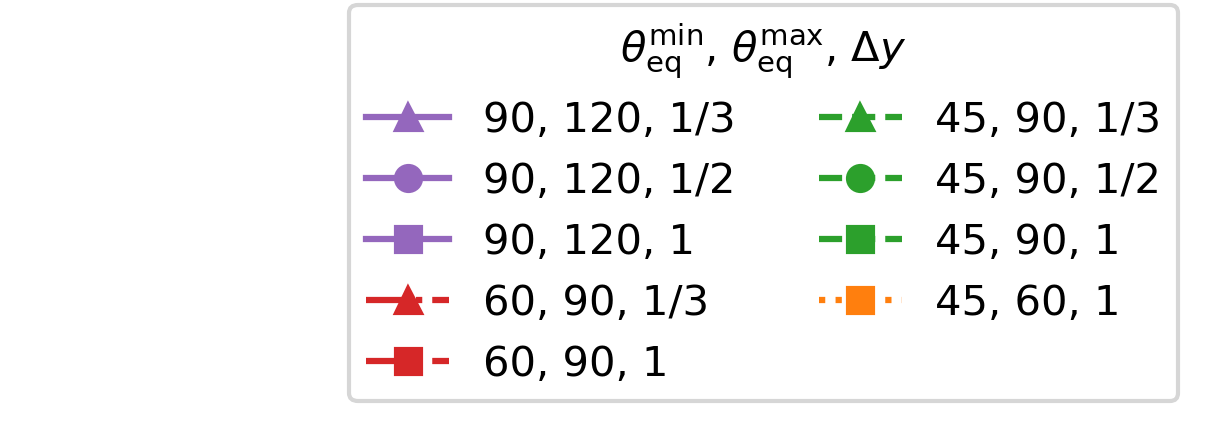}
  \begin{overpic}[width=\linewidth]{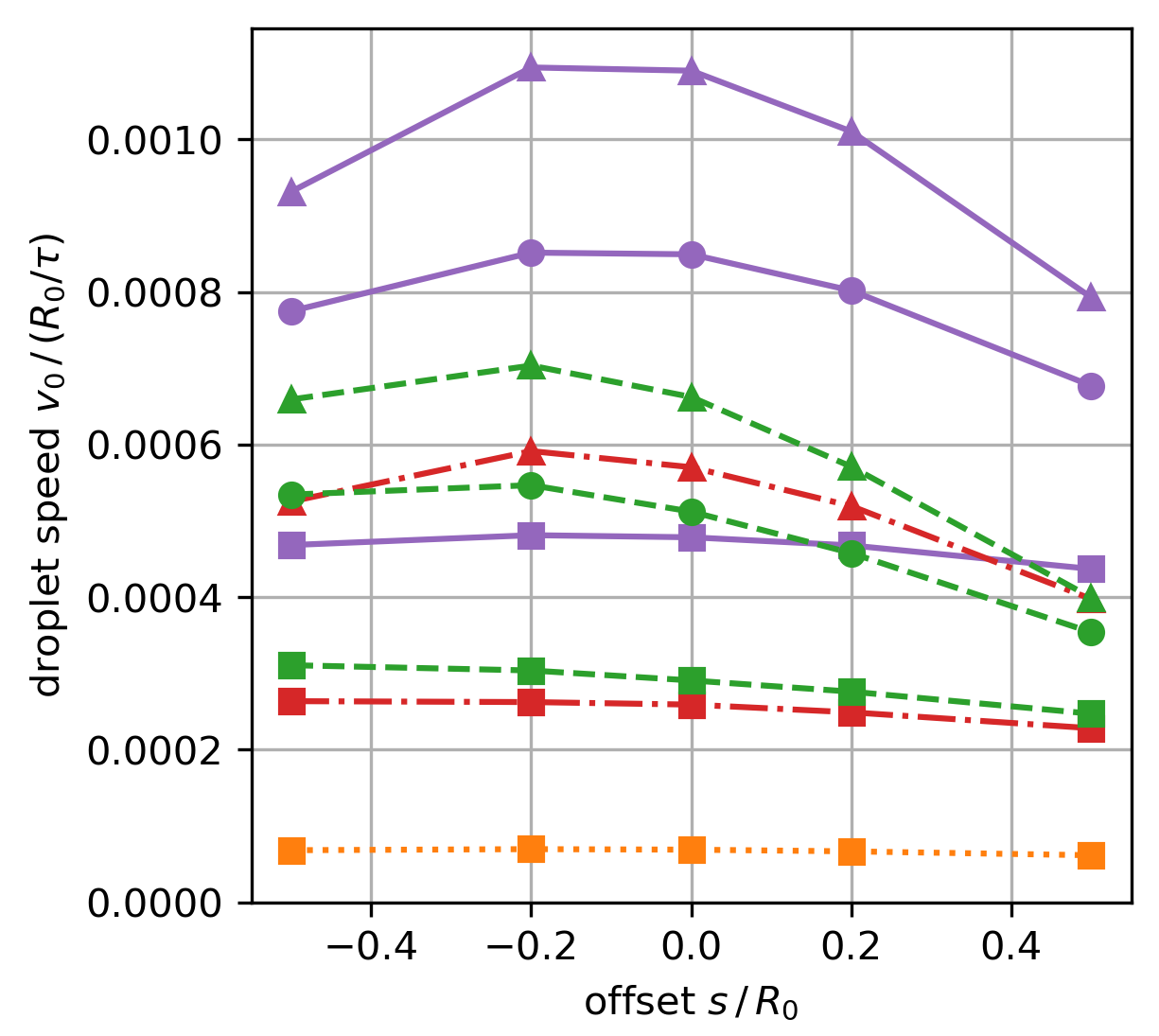}
  \put(7,84){\Large\textsf{(a)}}
  \end{overpic}
  \begin{overpic}[width=0.93\linewidth]{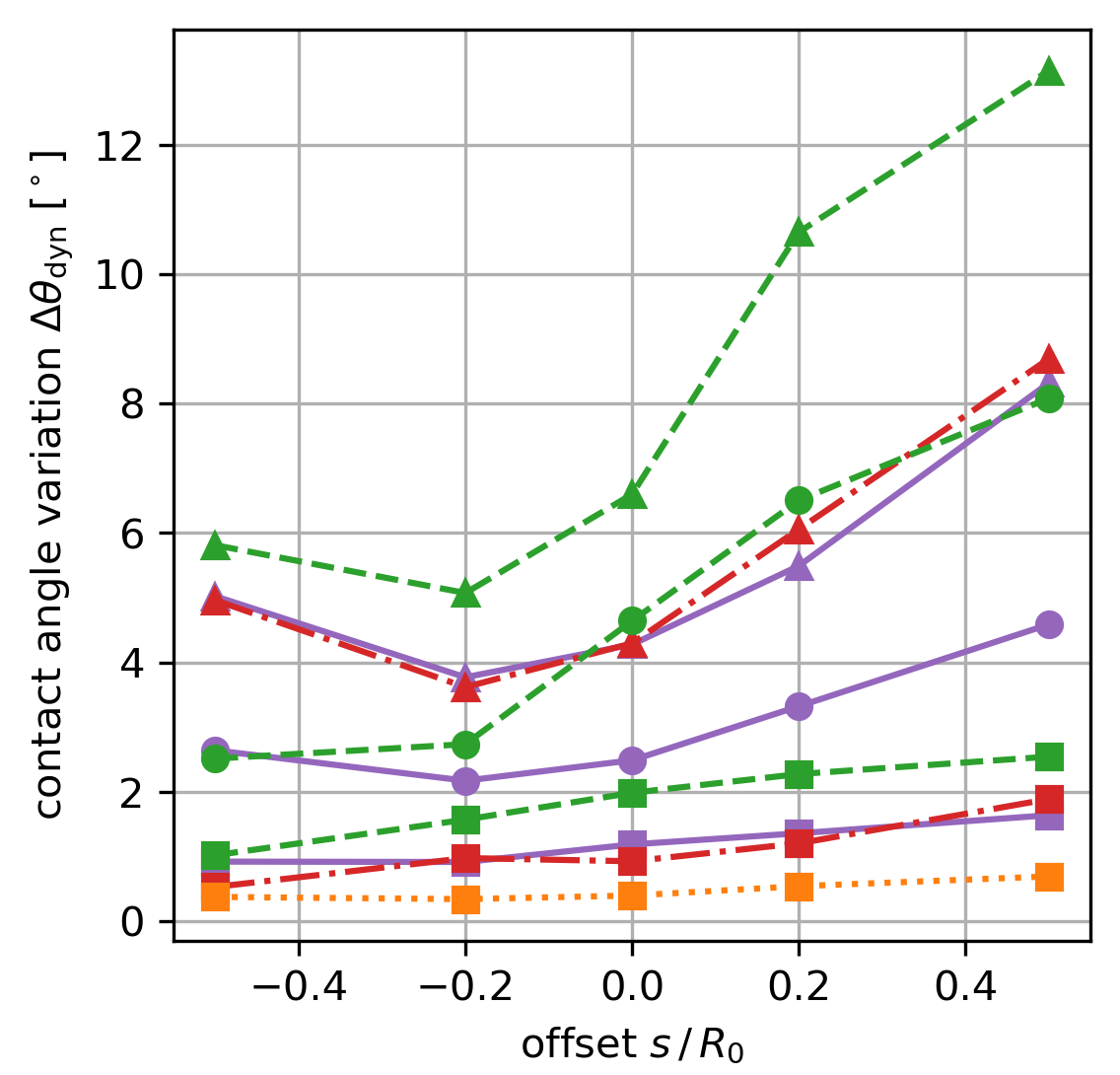}
  \put(0,90){\Large\textsf{(b)}}
  \put(17,53){\includegraphics[width=1.3in]{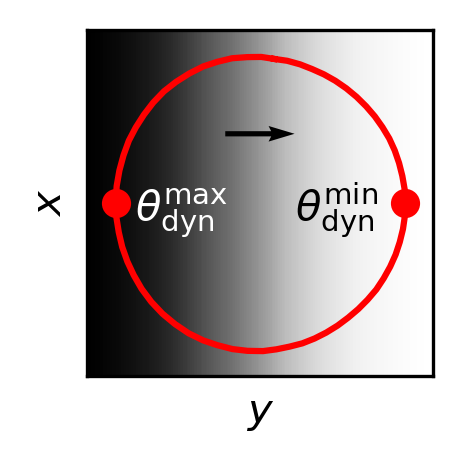}}
  \end{overpic}
 \caption{%
Steady-state (a) speed~$v_0$ and (b) maximal contact-angle difference~$\Delta \theta_\mathrm{dyn}$ plotted versus offset~$s$ for various feedback-driven droplets of a $90\,\%$ glycerol/$10\,\%$ water mixture in response to wettability step profiles characterized by different parameter sets (see legend).
Lines between symbols are added to guide the eye.
Inset in (b): Example for a contact line (red) surfing on a wettability step (background shading) with markers for the locations of
$\theta_\mathrm{dyn}^\mathrm{min}$ and  $\theta_\mathrm{dyn}^\mathrm{max}$ (circles).
}
 \label{fig_feedback}
\end{figure}

The steady-state droplet speed~$v_0$ depends on the various parameters of the wettability profile.
In Fig.~\ref{fig_feedback}(a) we plot the droplet speed versus the offset $s$ for several parameter sets structured by (i)~the limiting equilibrium contact angles on both sides of the wettability step and (ii)~the step width~$\varDelta y$.

First of all, we observe that $v_0$ increases overall with decreasing $\varDelta y$ which is consistent with the expectation that
steepness of the wettability step drives the droplet forward.
Furthermore, as $\varDelta y$ decreases for the same combinations of contact angles, an optimal offset $s$ becomes prominent close to $s=-0.2\,R_0$ (circle and triangle symbols in Fig.~\ref{fig_feedback}(a)), meaning the droplet lags behind the step and is pushed forward.
Since the wettability gradient is largest at $s=0$, we conclude that the steepness of the wettability step does not solely determine droplet speed.
Rather, there is a competing influence from the fact that the equilibrium contact angles~$\theta_\mathrm{eq}^\mathrm{max}$ and $\theta_\mathrm{eq}^\mathrm{min}$ contribute nonlinearly to droplet speed via the contact line.
According to the Cox-Voinov law the speed of the contact line depends on the difference of the \emph{cubes} of $\theta_\mathrm{dyn}$ and $\theta_\mathrm{eq}$.
Thus, $v_0$ increases with $\theta_\mathrm{eq}$ even when the difference $\theta_\mathrm{dyn}-\theta_\mathrm{eq}$ remains the same.
As a result, the droplet speed increases for larger $\theta_\mathrm{eq}$, which the droplet encounters if it lags behind at $s <0$.

This is nicely visible in Fig.~\ref{fig_feedback}(a), when we compare the $60^\circ$-$90^\circ$ combination in equilibrium contact angles (red symbols/dash-dotted lines) to the $90^\circ$-$120^\circ$ combination (purple symbols/full line).
Even though the difference between the limiting angles is $30^\circ$ in both cases, the droplet speed is larger for the larger equilibrium contact angles.

\subsection{Deformation}
Since the droplets are placed on a non-uniform wettability pattern, their shapes adapt and become asymmetric.
There is a competition between the Laplace pressure depending on the local curvature of the free surface, which drives the droplet toward a uniform curvature, and the forces acting on the contact line, which drive the local contact angle toward its equilibrium value.
Therefore, the free surface will deviate from a spherical cap.
The result of these competing forces can be observed in the variation of the dynamic contact angle~$\theta_\mathrm{dyn}$ along the contact line in steady state.
To quantify this variation, we plot in Fig.~\ref{fig_feedback}(b) the difference of the largest and smallest contact angle, $\Delta \theta_\mathrm{dyn} = \theta_\mathrm{dyn}^\mathrm{max} - \theta_\mathrm{dyn}^\mathrm{min}$, versus the offset $s$ for the different parameter sets.
Simply put, a larger difference $\Delta \theta_\mathrm{dyn}$ indicates a more asymmetric droplet shape.
Note that $\theta_\mathrm{dyn}^\mathrm{min}$ and $\theta_\mathrm{dyn}^\mathrm{max}$ are realized at the front and the back of the droplet w.r.t.~to its direction of motion as indicated in the inset of Fig.~\ref{fig_feedback}(b), where the equilibrium contact angles are extremal.
At these locations the contact line velocity $v_\mathrm{contact}$ must equal the droplet speed $v_0$ in steady state.
So the difference between the extrema of $\theta_\mathrm{dyn}$ and $\theta_\mathrm{eq}$ along the contact line
determines $v_0$.

We observe that $\Delta \theta_\mathrm{dyn}$ is primarily determined by large variations in $\theta_\mathrm{eq}$, which makes sense.
For example, the wettability profiles with $\theta_\mathrm{eq}^\mathrm{min}=45^\circ$ and $\theta_\mathrm{eq}^\mathrm{max}=90^\circ$ [green symbols/dashed lines in Fig.~\ref{fig_feedback}(b)] cause larger
variations~$ \Delta \theta_\mathrm{dyn}$ than step profiles with $\theta_\mathrm{eq}^\mathrm{min}=90^\circ$ and
$\theta_\mathrm{eq}^\mathrm{max}=120^\circ$ [purple symbols/solid lines in Fig.~\ref{fig_feedback}(b)] for otherwise
same parameters.

In addition, for $\varDelta y /R_0=1/3$ and $1/2$, $\Delta \theta_\mathrm{dyn}$ shows a strong dependence on the offset $s$.
Interestingly, for these cases the droplet asymmetry decreases notably as $s$ approaches the speed optimimum~$s=-0.2\, R_0$.
Apparently, a more symmetric droplet has larger differences between $\theta_\mathrm{dyn}$ and $\theta_\mathrm{eq}$ along its contact line and therefore achieves higher speeds~$v_0$.

In \emph{summary}, we observe droplets settle into a rolling motion when they are a placed on a wettability step profile with a constant offset to the location of steepest wettability gradient.
Their limiting speed increases with the steepness of the gradient.
For large steepness there is an optimal offset so that the droplets are pushed forward by regions of lower overall wettability.
Their shapes become more asymmetric in response to steeper gradients, while the optimal offset for speed is correlated with a more symmetric shape.

\section{Wettability step profile with constant driving}
\label{sec_constant_driving}

\subsection{Setup}
As an alternative to the feedback method described above, we study the simpler setup  where the wettability step of
Eq.~(\ref{eq_gradient}) moves with constant speed.
Here, two scenarios are possible:
If the droplet can keep up with the moving wettability profile, it will \emph{surf} on the pattern, otherwise it will fall behind, stop moving, and eventually adapt its shape to the lower wettability of the substrate.

To gain insight into these scenarios, we re-examine one particular pattern from the previous part: a wettability step with $\theta_\mathrm{eq}^\mathrm{min}=90^\circ$ and $\theta_\mathrm{eq}^\mathrm{max}=120^\circ$, and $\varDelta y=1/3$.
We place a glycerol droplet with parameters according to row~1 of Table~\ref{tab_parameters} well before the step and, unlike before, let the step approach the droplet with a constant speed~$v_\mathrm{s}$ independent of the position of the droplet.

\subsection{Long-time dynamics}
\label{sec_surfing_speed}

\begin{figure}
 \includegraphics[width=\linewidth]{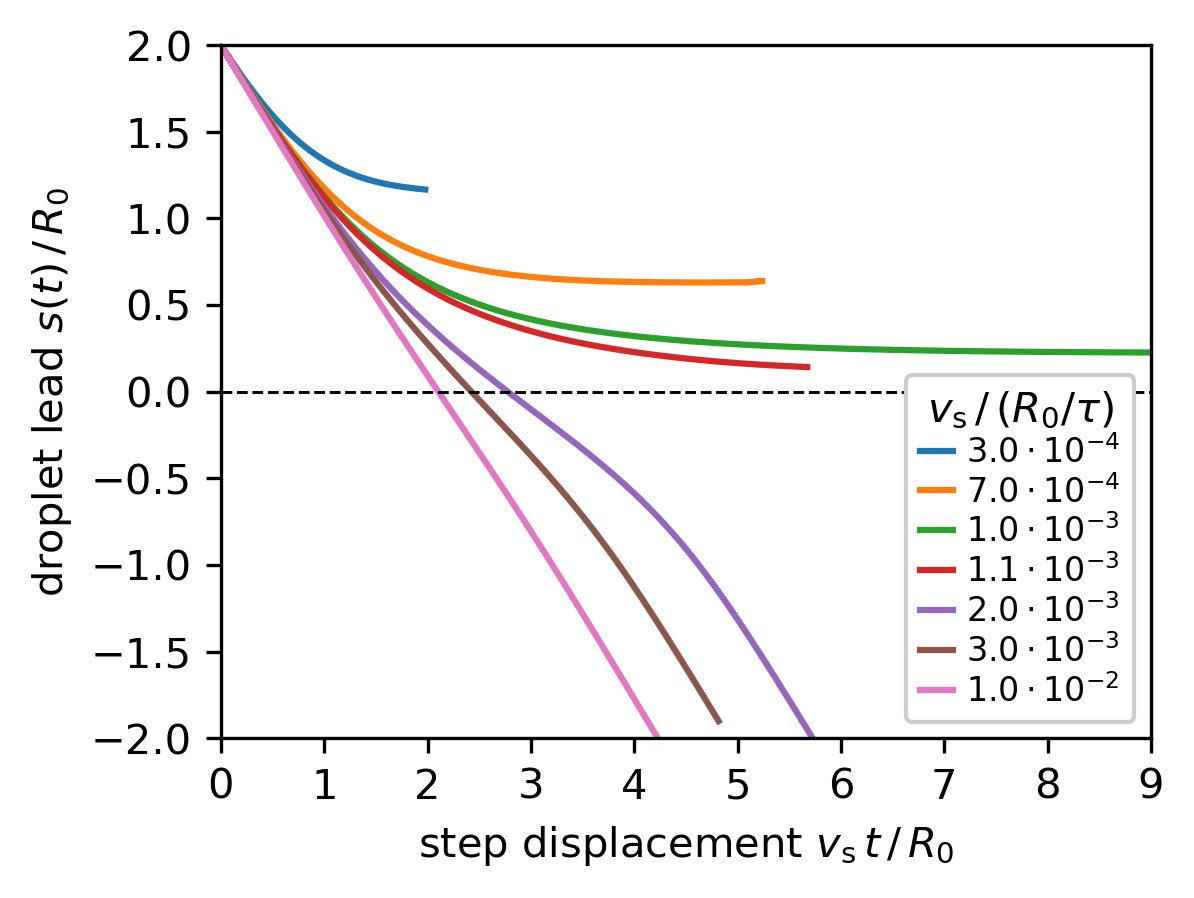}
 \caption{%
 Droplet lead~$s$ with initial value $s_0=2\,R_0$ plotted versus step displacement~$v_\mathrm{s}t$.
 The wettability step is characterized by $\theta_\mathrm{eq}^\mathrm{min}=90^\circ$, $\theta_\mathrm{eq}^\mathrm{max}=120^\circ$, and $\varDelta y = 1/3$ and for the droplet glycerol parameters (row~1 in Table~\ref{tab_parameters}) are taken.
 Solid lines indicate trajectories for various $v_\mathrm{s}$ as indicated in the legend.
 }
 \label{fig_constant_progress}
\end{figure}
We first position the droplet at $t=0$ with an initial lead~$s_0=2\,R_0$ before the steepest point of the step, which means the droplet is on the more wettable side of the step.
This gives the droplet some time to adapt to a surfing shape under the influence of the approaching step.
We then observe how the droplet lead,
\begin{equation}
 s(t)=y_\mathrm{c}(t) - v_\mathrm{s}t + s_0 \, ,
\end{equation}
evolves as step and droplet advance.
Here, $y_\mathrm{c}(t)$ is the distance traveled by the droplet.
Thus, in steady state, if the droplet can follow the wettability step, the droplet lead approaches a constant value.

In Fig.~\ref{fig_constant_progress} we display trajectories leading to either droplets surfing on the wettability step or droplets that cannot follow.
Above a step speed of roughly $v_\mathrm{s}=1.1 \cdot 10^{-3}\,R_0/\tau$ droplets cannot keep up with the step meaning their droplet lead tends to minus infinity.
Below, droplets match the step speed and surf, meaning their droplet lead approaches a constant value.
The transition between both scenarios occurs close to the maximum droplet speed of $1.1 \cdot 10^{-3}\,R_0 / \tau$, which we observed using the feedback mechanism in Sec.~\ref{sec_feedback_speed}, Fig.~\ref{fig_feedback}(a).
Furthermore, we note that the droplet lead of $0.24\,R_0$ observed for $v_\mathrm{s}=10^{-3}\,R_0/\tau$ is close to the
lead $0.2\,R_0$, which occured for a similar speed of $v_0=1.01\cdot10^{-3}\,R_0/\tau$ in Sec.~\ref{sec_feedback_speed} [purple triangles in Fig.~\ref{fig_feedback}(a)].

So far, we can draw two conclusions from our investigations in comparison to the feedback studies of the previous section.
First,  for a given step speed a droplet sitting ahead of the approaching wettability step assumes
the offset~$s$ given in the upper curve of Fig.~\ref{fig_feedback}(a) by the branch to the right of the maximum.
Second, droplets cannot follow wettability steps with larger speeds.

\subsection{Preferred surfing state}

\begin{figure}
 \includegraphics[width=\linewidth]{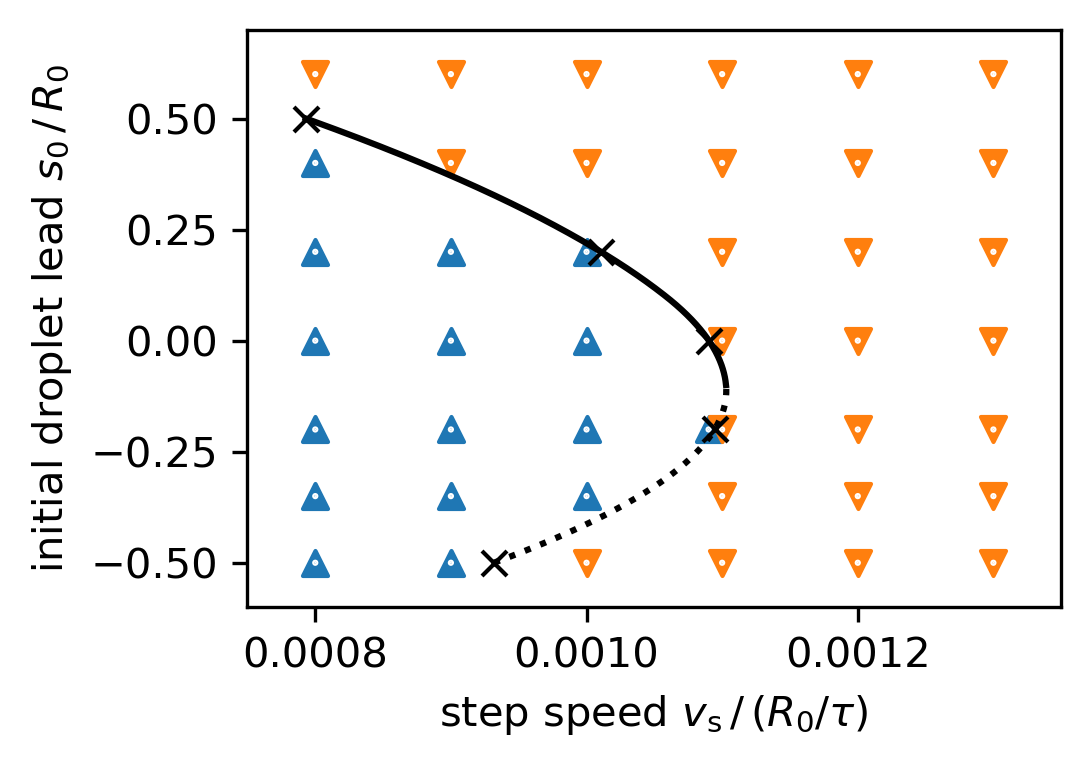}
 \caption{
Moving direction of the droplet plotted in a diagram versus initial lead~$s_0$ and step speed~$v_\mathrm{s}$.
Blue and yellow triangles indicate decreasing or increasing droplet lead, respectively.
The black solid-dashed curve corresponds to the steady-state motion from the feedback study in Sect.~\ref{sec_feedback} [\emph{cf.} the purple triangles in the upper curve of Fig.~\ref{fig_feedback}(a)].
Same parameters as in Fig.~\ref{fig_constant_progress} are used.
}
 \label{fig_Stability}
\end{figure}

How does the final surfing state of the droplet depend on the initial droplet lead $s_0$?
To answer this question, we placed a droplet with spherical cap shape and contact angle $\theta_\mathrm{dyn}=90^\circ$ at different positions relative to the wettability step, which moved with different velocities.
The result is presented in the state diagram $s_0$ \emph{vs.} $v_s$ of Fig.~\ref{fig_Stability}.
The black solid-dashed line indicates the result from the feedback analysis in Fig.~\ref{fig_feedback}(a).
The yellow triangles mean that the droplet lead decreases.
For $v_\mathrm{s} \leq 1.1\,R_0 / \tau$, when the lead reaches the upper branch of the black curve, the droplet is taken along by the step and moves with constant speed.
For $v_\mathrm{s} > 1.1\,R_0 / \tau$ the droplet cannot follow the wettability step and is left behind
(see movie~M03 in ESI$^\dag$).
The blue triangles inside the black curve indicate droplets that initially move faster than the wettability step.
The droplet lead $s$ increases until it hits again the solid black line and the droplet again moves with constant
speed $v_s$ (see movie~M04 in ESI$^\dag$).

So the droplet speed determined in our feedback analysis as a function of the offset or droplet lead~$s$ gives exactly the outcome when we use a constant step speed.
However, the black curve in Fig.~\ref{fig_Stability} has a stable branch (solid line) indicating that the droplet will mostly move in front of the wettability step, while the positions given by the dashed line are unstable.
This behavior is reminiscent of a \emph{saddle-node bifurcation} near $v_\mathrm{s} \approx 1.1\,R_0 / \tau$.
Our results also show that for droplets on wettability steps feedback
has a stabilizing effect on an otherwise unstable state (see, \emph{e.g.}, Ref.~\cite{pyragas_continuous_1992}).

The (in)stability of either branch is intuitively explained by the steepness and local curvature of the sigmoid curve, which we use for the wettability profile (see Fig.~\ref{fig_feedback_schematic}):
By definition, a leading droplet is on the \emph{convex} part of the sigmoid which means, as it falls behind it is exposed to an increasing gradient which speeds it up again.
Conversely, a lagging droplet is on the \emph{concave} part of the sigmoid and as it falls behind it is exposed to a decreasing gradient which slows it down further.
Only close to the turning point also lagging positions are stable.
We explain this by the fact that the droplet is an
extended object.
The unstable and stable positions meet at the bifurcation point, which is near the steepest part of the sigmoid.
There, the droplet is exposed to the largest possible gradient and therefore achieves its maximum speed.

\section{Conclusions}
\label{sec_conclusions}
In this article we have applied the boundary element method and developed a numerical scheme to simulate droplets on  substrates with switchable and non-uniform wettability realized, for example, by light-switchable surfaces.
A strong emphasis was put on developing a stable grid on the droplet surface.
We are able to simulate droplets on the whole range of possible contact angles and, in particular, investigated speed and deformation of droplets under the influence of moving wettability steps.
A crucial ingredient for simulating the dynamics of droplets is the Cox-Voinov law, which governs the motion of the three-phase contact line.
In addition, using the boundary element method allows us to calculate flow fields inside the droplet and not just on its surface and thereby provides a further understanding of the droplet motion.

We first applied our method to sessile droplets and exposed them to a spatially uniform but instantaneous change in wettability.
We carefully compared our findings to experimental results from Ref.~\cite{deruijter_contact_1997} and found quantitative agreement.
The spherical cap model, where the dynamics is governed by the dynamics of the contact line via the Cox-Voinov law, shows small but noticeable differences.

We then applied our numerical scheme to droplets surfing on moving wettability steps.
First, using a feedback loop we kept the center of the wettability step at a constant distance or offset from the droplet center.
For a range of offsets and step widths, this gives rise to droplets surfing at constant speed on the wettability step into the direction of higher wettability.
Inside the droplet a single vortex forms such that the droplet rolls on the substrate.
For shallow wettability steps the resulting droplet speed hardly depends on the droplet offset in the range investigated.
However, for steeper steps the speed develops a maximum at offsets behind the steepest point of the step so that the droplet experiences regions of lower overall wettability.
Additionally, while droplets surfing on steep steps are less symmetric compared to more shallow steps, droplets surfing with the maximum speed are more symmetric compared to
slower droplets on the same wettability profile.

In a second study we moved the wettability step at a constant speed with various initial offsets between the step and droplet center.
Beyond the maximal droplet speed from the feedback study, steady surfing is not possible.
The droplet cannot follow the wettability step and is left behind.
Below the maximal surfing speed, droplets always settle to the same offset as observed in the feedback study.
However, only the leading positions where the droplet is pushed forward are stable while the lagging positions are unstable in the absence of feedback.
We explain this by the curvature of the smooth wettability step:
In the leading position the convex variation of the wettability step generates an effective restoring force when the droplet is left behind, while in the lagging position the concave variation cannot stabilize the surfing droplet against small displacements away from the step center.

Our findings demonstrate how a dynamic wettability profile, for example, controlled by structured light can be used to continuously drive droplets forward.
We have investigated here the important parameters to optimize the speed of droplets w.r.t.~the shape of the wettability profile and its driving speed.
Because a single wettability step can be set up to move droplets in any direction, in principle, we provide here a tool to steer droplets along arbitrary paths and with variable speed on a light-switchable substrate.
In the future, we plan to investigate droplet dynamics in more complex spatio-temporal wettability patterns motivated by the possibilities of structured light~\cite{rubinsztein_roadmap_2017}.

\section*{Conflicts of interest}
There are no conflicts to declare.

\section*{Acknowledgements}
We thank T.~Mallouk, D.~Peschka, and J.~Simmchen for helpful discussions and acknowledge financial support from DFG (German Research Foundation) \emph{via} Collaborative Research Center~910.

\appendix
\section{Mesh degradation}
\label{sec_mesh_benchmark}

\begin{figure}
 \centering
 \includegraphics[width=\linewidth]{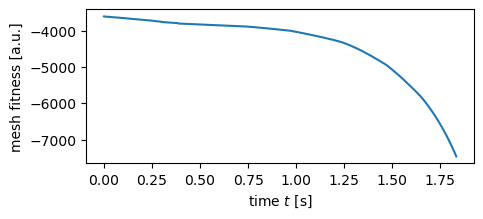}
 \caption{
 The mesh fitness~$M$ for a sessile glycerol droplet decreases over time in response to a wettability change at $t=0\,\mathrm{s}$.}
 \label{fig_mesh_fitness_glycerin}
\end{figure}
Over the course of a simulation, mesh fitness~$M$ (see Ref.~\cite{zinchenko_emulsion_2013} for definition) degrades due to physically necessary changes in the mesh shape.
In Fig.~\ref{fig_mesh_fitness_glycerin} we give an example of mesh degradation during the initial relaxation of a droplet on a uniform substrate presented in Sec.~\ref{sec_validation}.
Notably, $M$ initially decreases roughly linearly while the droplet changes shape, as visible in Fig.~\ref{fig_glycerin}.
However, $M$ decreases further thereafter which implies that, even in the absence of further deformations, numerical error still compounds.
Small numerical errors have a compounding effect on $M$ because $M$ is a nonlinear function of vertex positions,
and therefore displacements of similar magnitude have a small effect on the initial (effectively optimal) initial mesh for $t<1\,\mathrm{s}$ and a disproportionally larger effect on the already deformed mesh for $t>1\,\mathrm{s}$.

\section{Parameters used for movies in ESI}
\label{appendix_esi}
All movies visualize simulations with material parameters from row~1 of Table~\ref{tab_parameters}.
Movie~M01 corresponds to the last example provided in Sec.~\ref{sec_validation} and Fig.~\ref{fig_60_120_relaxation}, which means it shows an instantaneous change in uniform wettability from $\theta_\mathrm{eq}=60^\circ$ to $120^\circ$.
Movie~M02 corresponds to the purple triangle at offset~$-0.2\,R_0$ in Fig.~\ref{fig_feedback}(a).
Movie~M03 shows a droplet driven by a wettability step from $90^\circ$ to $120^\circ$ with steepness~$\varDelta y=1/3$, step speed of $v_\mathrm{s}=2\cdot10^{-3}\,R_0/\tau$, and an initial lead of $s_0=2\,R_0$.
Movie~M04 corresponds to the same parameters, except~$v_\mathrm{s}=1\cdot10^{-3}\,R_0/\tau$.
Movie~M05 corresponds to the parameters given in Fig.~\ref{fig_feedback_flow}.
Note that the movies each run at different speeds.


\printbibliography[heading=bibintoc]

@article{palagi_structured_2016,
    author={Palagi, Stefano
    and Mark, Andrew G.
    and Reigh, Shang Yik
    and Melde, Kai
    and Qiu, Tian
    and Zeng, Hao
    and Parmeggiani, Camilla
    and Martella, Daniele
    and Sanchez-Castillo, Alberto
    and Kapernaum, Nadia
    and Giesselmann, Frank
    and Wiersma, Diederik S.
    and Lauga, Eric
    and Fischer, Peer},
    title={Structured light enables biomimetic swimming and versatile locomotion of photoresponsive soft microrobots},
    journal={Nat. Mater.},
    year={2016},
    volume={15},
    number={6},
    pages={647-653},
    issn={1476-4660},
    doi={10.1038/nmat4569},
    url = {https://www.repository.cam.ac.uk/handle/1810/253886},
    url+an = {=openaccess}
}

@article{chaudhury_how_1992,
    title   = {How to Make Water Run Uphill},
    author  = {Manoj K. Chaudhury and George M. Whitesides},
    journal = {Science},
    volume  = {256},
    pages   = {1539},
    year    = {1992},
    doi     = {10.1126/science.256.5063.1539}
}

@article{ichimura_light_2000,
    title   = {Light-Driven Motion of Liquids on a Photoresponsive Surface},
    author  = {Kunihiro Ichimura and Sangkeun Oh and Masaru Nakagawa},
    journal = {Science},
    volume  = {288},
    pages   = {1624},
    year    = {2000},
    doi     = {10.1126/science.288.5471.1624},
    eprinttype = {semanticscholar},
    eprint = {25350736},
    eprint+an = {=openaccess}
}

@article{glasner_boundary_2005,
    title   = {A boundary integral formulation of quasi-steady fluid wetting},
    author  = {K. B. Glasner},
    journal = {J. Comp. Phys.},
    volume  = {207},
    pages   = {529},
    year    = {2005},
    doi     = {10.1016/j.jcp.2005.01.022}
}

@article{alinovi_boundary_2017,
    title   = {A boundary element method for Stokes flows with interfaces},
    author  = {Edoardo Alinovi and Alessandro Bottaro},
    journal = {J. Comp. Phys.},
    volume  = {356},
    pages   = {261},
    year    = {2017},
    doi     = {10.1016/j.jcp.2017.12.004}
}

@article{lim_photoreversibly_2006,
    title   = {Photoreversibly Switchable Superhydrophobic Surface with Erasable and Rewritable Pattern},
    author  = {Ho Sun Lim and Joong Tark Han and Donghoon Kwak and Meihua Jin and Kilwon Cho},
    journal = {J. Am. Chem. Soc.},
    volume  = {128},
    pages   = {14458},
    year    = {2006},
    doi     = {10.1021/ja0655901}
}

@article{baigl_photo_2012,
    title   = {Photo-actuation of liquids for light-driven microfluidics: state of the art and perspectives},
    author  = {Damien Baigl},
    journal = {Lab Chip},
    volume  = {12},
    pages   = {3637},
    year    = {2012},
    doi     = {10.1039/c2lc40596b}
}

@article{wang_photoresponsive_2007,
    title   = {Photoresponsive surfaces with controllable wettability},
    author  = {Shutao Wang and Yanlin Song and Lei Jiang},
    journal = {J. Photochem. Photobiol. C: Photochem. Rev.},
    volume  = {8},
    pages   = {18},
    year    = {2007},
    doi     = {10.1016/j.jphotochemrev.2007.03.001}
}

@article{blossey_self_2003,
    title   = {Self-cleaning surfaces — virtual realities},
    author  = {Ralf Blossey},
    journal = {Nat. Mater.},
    volume  = {2},
    pages   = {301},
    year    = {2003},
    doi     = {10.1038/nmat856}
}

@article{schmitt_marangoni_2016,
    title   = {Marangoni flow at droplet interfaces: Three-dimensional solution and applications},
    author  = {M. Schmitt and H. Stark},
    journal = {Phys. Fluids},
    volume  = {28},
    pages   = {012106},
    year    = {2016},
    doi     = {10.1063/1.4939212},
    eprinttype = {arxiv},
    eprint = {1512.05721}
}

@article{seki_photo_2005,
    title   = {Photo-driven directional motion of droplets on the surface of a liquid crystal doped with photochromic azobenzene: theory},
    author  = {Kazuhiko Seki and M Tachiya},
    journal = {J. Phys.: Condens. Matter},
    volume  = {17},
    pages   = {S4229},
    year    = {2005},
    doi     = {10.1088/0953-8984/17/49/016}
}

@article{varanakkottu_light_2016,
    title   = {Light-Directed Particle Patterning by Evaporative Optical Marangoni Assembly},
    author  = {Subramanyan Namboodiri Varanakkottu and Manos Anyfantakis and Mathieu Morel and Sergii Rudiuk and Damien Baigl},
    journal = {Nano Lett.},
    volume  = {16},
    pages   = {644},
    year    = {2016},
    doi     = {10.1021/acs.nanolett.5b04377}
}

@article{grawitter_feedback_2018,
    journal = {Soft Matter},
    title = {Feedback control of photoresponsive fluid interfaces},
    year = {2018},
    doi = {10.1039/c7sm02101a},
    volume = {14},
    author = {Grawitter, Josua and Stark, Holger},
    pages = {1856},
    doi+an  = {=openaccess}
}

@article{xiao_moving_2018,
    title = {Moving Droplets in 3D Using Light},
    author = {Yang Xiao and Sara Zarghami and Klaudia Wagner and Pawel Wagner  Keith C. Gordon and Larisa Florea Dermot Diamond and David L. Officer},
    journal = {Adv. Mater.},
    volume = {30},
    pages = {1801821},
    year = {2018},
    doi = {10.1002/adma.201801821}
}

@article{kaspar_confinement_2016,
    title = {Confinement of water droplets on rectangular micro/nano-arrayed surfaces},
    author = {Kašpar, Ondřej and Zhang, Hailong and Tokárová, Viola and Boysen, Reinhard I. and Suñé, Gemma Rius and Borrise, Xavier and Perez-Murano, Francesco and Hearn, Milton T. W. and Nicolau, Dan V.},
    journal = {Lab Chip},
    volume = {16},
    pages = {2487},
    year = {2016},
    doi = {10.1039/c6lc00622a}
}

@article{pozrikidis_interfacial_2001,
    title = {Interfacial Dynamics for Stokes Flow},
    author = {C. Pozrikidis},
    journal = {J. Comp. Phys.},
    volume = {169},
    pages = {250},
    year = {2001},
    doi = {10.1006/jcph.2000.6582}
}

@article{edalatpour_managing_2018,
    title = {Managing water on heat transfer surfaces: A critical review of techniques to modify surface wettability for applications with condensation or evaporation},
    author = {M. Edalatpour and L. Liu and A.M. Jacobi and K.F. Eid and A.D. Sommers},
    journal = {Appl. Energy},
    volume = {222},
    pages = {967},
    year = {2018},
    doi = {10.1016/j.apenergy.2018.03.178}
}

@article{darhuber_principles_2005,
    title = {Principles of microfluidic actuation by modulation of surface stresses},
    author = {Anton A. Darhuber and Sandra M. Troian},
    journal = {Ann. Rev. Fluid Mech.},
    volume = {37},
    pages = {425},
    year = {2005},
    doi = {10.1146/annurev.fluid.36.050802.122052},
    eprinttype = {caltech},
    eprint = {DARarfm05},
    eprint+an = {=openaccess}
}

@book{kim_microhydrodynamics_2005,
    title = {Microhydrodynamics},
    author = {Sangtae Kim and Sepp J. Karrila},
    year = {2005},
    isbn = {0486442195},
    publisher = {Dover Publications},
    address = {Mineola/NY}
}

@article{li_evaporation_2013,
    title = {Evaporation Stains: Suppressing the Coffee-Ring Effect by Contact Angle Hysteresis},
    author = {Li, Yuehfeng and Sheng, Yujane and Tsao, Hengkwong},
    journal = {Langmuir},
    volume = {29},
    pages = {7802},
    year = {2013},
    doi = {10.1021/la400948e}
}

@article{bonn_wetting_2009,
    title = {Wetting and spreading},
    author = {Daniel Bonn and Jens Eggers and Joseph Indekeu and Jacques Meunier and Etienne Rolley},
    journal = {Rev. Mod. Phys.},
    volume = {81},
    pages = {739},
    year = {2009},
    doi = {10.1103/RevModPhys.81.739}
}

@book{pozrikidis_practical_2002,
    title = {A practical guide to boundary element methods with software library BEMLIB},
    author = {Costas Pozrikidis},
    publisher = {CRC Press},
    address = {Boca Raton/FL},
    year = {2002},
    isbn = {1584883235}
}

@article{mugele_electrowetting_2005,
    title = {Electrowetting: a convenient way to switchable wettability patterns},
    author = {F Mugele and A Klingner and J Buehrle and D Steinhauser and S Herminghaus},
    journal = {J. Phys.: Condens. Matter},
    volume = {17},
    pages = {S559},
    year = {2005},
    doi = {10.1088/0953-8984/17/9/016}
}

@book{pozrikidis_boundary_1992,
    title = {Boundary integral and singularity methods for linearized viscous flow},
    author = {C. Pozrikidis},
    publisher = {Cambridge University Press},
    address = {Cambridge},
    isbn = {9780521406932},
    year = {1992}
}

@article{chevallier_pumping_2011,
    author = {Chevallier, E. and Mamane, A. and Stone, H. A. and Tribet, C. and Lequeux, F. and Monteux, C.},
    title = {Pumping-out photo-surfactants from an air–water interface using light},
    journal = {Soft Matter},
    year = {2011},
    volume = {7},
    issue = {17},
    pages = {7866-7874},
    publisher = {The Royal Society of Chemistry},
    doi = {10.1039/C1SM05378G},
}

@article{ledesma_theory_2013,
    title = {Theory of Wetting-Induced Fluid Entrainment by Advancing Contact Lines on Dry Surfaces},
    author = {R. Ledesma-Aguilar and A. Hern\'andez-Machado and I. Pagonabarraga},
    journal = {Phys. Rev. Lett.},
    volume = {110},
    pages = {264502},
    year = {2013},
    doi = {10.1103/PhysRevLett.110.264502},
    eprinttype = {hdl},
    eprint = {2445/49243},
    eprint+an = {=openaccess}
}

@article{eral_contact_2013,
    title = {Contact angle hysteresis: a review of fundamentals and applications},
    author = {H. B. Eral and D. J. C. M. {'t Mannetje} and J. M. Oh},
    journal = {Colloid Polym. Sci.},
    volume = {291},
    pages = {247},
    year = {2013},
    doi = {10.1007/s00396-012-2796-6},
    eprinttype = {hdl},
    eprint = {1721.1/86387},
    eprint+an = {=openaccess}
}

@article{voinov_hydrodynamics_1976,
    title = {Hydrodynamics of Wetting},
    author = {O. V. Voinov},
    journal = {Fluid Dyn.},
    volume = {11},
    pages = {714},
    year = {1976},
    doi = {10.1007/BF01012963}
}

@article{cox_dynamics_1986,
    title = {The dynamics of the spreading of liquids on a solid surface},
    author = {R. G. Cox},
    journal = {J. Fluid Mech.},
    volume = {168},
    pages = {169},
    year = {1986},
    doi = {10.1017/S0022112086000332}
}

@article{snoeijer_moving_2013,
    title = {Moving Contact Lines: Scales, Regimes, and Dynamical ransitions},
    author = {Snoeijer, Jacco H. and Andreotti, Bruno},
    journal = {Annu. Rev. Fluid Mech.},
    volume = {45},
    pages = {269},
    year = {2013},
    doi = {10.1146/annurev-fluid-011212-140734},
    eprinttype = {semanticscholar},
    eprint = {121130784},
    eprint+an = {=openaccess}
}

@article{pityuk_boundary_2018,
    title = {Boundary Element Modeling of Dynamics of a Bubble in Contact with a Solid Surface at Low Reynolds Numbers},
    author = {Yu. A. Pityuk and O. A. Abramova and N. A. Gumerov and I. Sh. Akhatov},
    journal = {Math. Models Comput. Simul.},
    volume = {10},
    pages = {209},
    year = {2018},
    doi = {10.1134/S2070048218020102}
}

@article{tsitouras_runge_2011,
    title = {Runge–Kutta pairs of order 5(4) satisfying only the first column simplifying assumption},
    author = {Tsitouras, {Ch.}},
    journal = {Comput. Math. Appl.},
    volume = {62},
    pages = {770},
    year = {2011},
    doi = {10.1016/j.camwa.2011.06.002}
}

@article{mcgraw_slip_2016,
    title = {Slip-mediated dewetting of polymer microdroplets},
    author = {Joshua D. McGraw and Tak Shing Chan and Simon Maurer and Thomas Salez and Michael Benzaquen and Elie Raphaël and Martin Brinkmann and Karin Jacobs},
    journal = {Proc. Natl. Acad. Sci. U.S.A.},
    volume = {113},
    pages = {1168},
    year = {2016},
    doi = {10.1073/pnas.1513565113},
    doi+an  = {=openaccess}
}

@article{bolanos_derivation_2017,
    title = {Derivation of the Navier slip and slip length for viscous flows over a rough boundary},
    author = {Silvia Jiménez Bolaños and Bogdan Vernescu},
    journal = {Phys. Fluids},
    volume = {29},
    pages = {057103},
    year = {2017},
    doi = {10.1063/1.4982899}
}

@article{guckenberger_bending_2016,
    title = {On the bending algorithms for soft objects in flows},
    author = {Achim Guckenberger and Marcel P. Schraml and Paul G. Chen and Marc Leonetti and Stephan Gekle},
    journal = {Comput. Phys. Commun.},
    volume = {207},
    pages = {1-23},
    year = {2016},
    doi = {10.1016/j.cpc.2016.04.018},
    eprinttype = {hal},
    eprint = {hal-01314722},
    eprint+an = {=openaccess}
}

@article{zinchenko_emulsion_2013,
    title = {Emulsion flow through a packed bed with multiple drop breakup},
    author = {Alexander Z. Zinchenko and Robert H. Davis},
    journal = {J. Fluid Mech.},
    volume = {725},
    pages = {611},
    year = {2013},
    doi = {10.1017/jfm.2013.197}
}

@article{deruijter_contact_1997,
    title = {Contact Angle Relaxation during the Spreading of Partially Wetting Drops},
    author = {de Ruijter, M. J. and De Coninck, J. and Blake, T. D. and Clarke, A. and Rankin, A.},
    journal = {Langmuir},
    volume = {13},
    pages = {7293},
    year = {1997},
    doi = {10.1021/la970825v}
}

@book{katsikadelis_boundary_2016,
    title = {The Boundary Element Method for Engineers and Scientists},
    author = {John T. Katsikadelis},
    publisher = {Elsevier},
    year = {2016},
    isbn = {978-0-12-804493-3},
    edition = {2}
}

@article{pyragas_continuous_1992,
    title = {Continuous control of chaos by self-controlling feedback},
    author = {Kestutis Pyragas},
    journal = {Phys. Lett. A},
    volume = {170},
    pages = {421},
    year = {1992},
    doi = {10.1016/0375-9601(92)90745-8}
}

@article{pirani_light_2016,
    title = {Light-Driven Reversible Shaping of Individual Azopolymeric Micro-Pillars},
    author = {Federica Pirani and Angelo Angelini and Francesca Frascella and Riccardo Rizzo and Serena Ricciardi and Emiliano Descrovi},
    journal = {Sci. Rep.},
    volume = {6},
    pages = {31702},
    year = {2016},
    doi = {10.1038/srep31702},
    doi+an  = {=openaccess}
}

@article{zhu_light_2020,
    title = {The light-driven macroscopic directional motion of a water droplet on an azobenzene–calix[4]arene modified surface},
    author = {Fei Zhu and  Shiliang Tan and  Manivannan Kalavathi Dhinakaran and Jing Cheng and Haibing Li},
    journal = {Chem. Comm.},
    volume = {56},
    pages = {10922},
    year = {2020},
    doi = {10.1039/D0CC00519C}
}

@article{gao_droplets_2018,
    title = {Droplets Manipulated on Photothermal Organogel Surfaces},
    author = {Chunlei Gao and
Lei Wang and
Yucai Lin and
Jitong Li and
Yufang Liu and
Xin Li and
Shile Feng and
Yongmei Zheng},
    journal = {Adv. Funct. Mater.},
    volume = {28},
    pages = {1803072},
    year = {2018},
    doi = {10.1002/adfm.201803072}
}

@article{savva_droplet_2019,
    title = {Droplet dynamics on chemically heterogeneous substrates},
    author = {Nikos Savva and Danny Groves and Serafim Kalliadasis},
    journal = {J. Fluid Mech.},
    volume = {859},
    pages = {321},
    year = {2019},
    doi = {10.1017/jfm.2018.758},
    url = {http://orca.cf.ac.uk/id/eprint/118322},
    url+an = {=openaccess}
}

@article{chan_morphological_2017,
    title = {Morphological evolution of microscopic dewetting droplets with slip},
    author = {Chan, Tak Shing and McGraw, Joshua D. and Salez, Thomas and Seemann, Ralf and Brinkmann, Martin},
    journal = {J. Fluid Mech.},
    volume = {828},
    pages = {271},
    year = {2017},
    doi = {10.1017/jfm.2017.515},
    eprinttype = {hdl},
    eprint = {2115/68436},
    eprint+an = {=openaccess}
}

@article{squires_microfluidics_2005,
    title = {Microfluidics:~Fluid physics at the nanoliter scale},
    author = {Todd M. Squires and Stephen R. Quake},
    journal = {Rev. Mod. Phys.},
    volume = {77},
    pages = {977},
    year = {2005},
    doi = {10.1103/RevModPhys.77.977},
    eprinttype = {caltech},
    eprint = {SQUrmp05},
    eprint+an = {=openaccess}
}

@article{rubinsztein_roadmap_2017,
    title = {Roadmap on structured light},
    author = {Halina Rubinsztein-Dunlop and Andrew Forbes and M V Berry and M R Dennis and David L Andrews and Masud Mansuripur and Cornelia Denz and Christina Alpmann and Peter Banzer and Thomas Bauer and Ebrahim Karimi and Lorenzo Marrucci and Miles Padgett and Monika Ritsch-Marte and Natalia M Litchinitser and Nicholas P Bigelow and C Rosales-Guzmán and A Belmonte and J P Torres and Tyler W Neely and Mark Baker and Reuven Gordon and Alexander B Stilgoe and Jacquiline Romero and Andrew G White and Robert Fickler and Alan E Willner and Guodong Xie and Benjamin McMorran and Andrew M Weiner},
    journal = {J. Opt.},
    volume = {19},
    pages = {013001},
    year = {2017},
    doi = {10.1088/2040-8978/19/1/013001},
    eprinttype = {hdl},
    eprint = {2117/107754},
    eprint+an = {=openaccess}
}
\end{document}